\documentclass[a4paper,11pt]{article}
\usepackage{jinstpub} 

\usepackage{amssymb}
\usepackage{graphicx}
\usepackage{caption}    
\usepackage{subcaption}
\usepackage{siunitx}
\usepackage{lineno}
\usepackage{tabularx}
\usepackage{array}
\usepackage{ragged2e}

\newcolumntype{C}{>{\centering\arraybackslash}X}

\usepackage{wasysym} 
\usepackage{xcolor}
\newcommand{\added}[1]{#1}                    

\title{\boldmath Performance Optimization and Characterization of 7-Pad Resistive PICOSEC Micromegas Detectors}

\author[a,1]{A. Kallitsopoulou,\note{Corresponding author}}
\author[a]{R. Aleksan}
\author[b]{S. Aune}
\author[d]{J. Bortfeldt}
\author[e]{F. Brunbauer}
\author[f,g]{M. Brunoldi}
\author[h]{J. Datta}
\author[a]{D. Desforge}
\author[i]{G. Fanourakis}
\author[f,g,2]{D. Fiorina\note{Now at Gran Sasso Science Institute, Viale F. Crispi, 7 67100 L'Aquila, Italy}}
\author[e,j]{K. J. Floethner}
\author[k]{M. Gallinaro}
\author[l]{F. Garcia}
\author[a]{I. Giomataris}
\author[m]{K. Gnanvo}
\author[a,3]{F.J. Iguaz\note{Now at SOLEIL Synchrotron, L'Orme des Merisiers, 91190 Saint Aubin, France}}
\author[e]{D. Janssens}
\author[a]{F. Jeanneau}
\author[n]{M. Kovacic}
\author[m]{B. Kross}
\author[a]{P. Legou}
\author[e]{M. Lisowska}
\author[o]{J. Liu}
\author[j,p]{M. Lupberger}
\author[b,c,4]{I. Maniatis\note{Now at Department of Particle Physics and Astronomy, Weizmann Institute of Science, Rehovot, Israel.}}
\author[m]{J. McKisson}
\author[o]{Y. Meng}
\author[e,p]{H. Muller}
\author[e]{E. Oliveri}
\author[e,q]{G. Orlandini}
\author[m]{A. Pandey}
\author[a]{T. Papaevangelou}
\author[r]{M. Pomorski}
\author[a]{E.Ferrer-Ribas}
\author[e]{L. Ropelewski}
\author[b,c]{D. Sampsonidis}
\author[e]{L. Scharenberg}
\author[e]{T. Schneider}
\author[r]{E. Scorsone}
\author[b,5]{L. Sohl\note{Now at TUV NORD EnSys GmbH \& Co. KG.}}
\author[e]{M. van Stenis}
\author[s]{Y. Tsipolitis}
\author[b,c]{S.E. Tzamarias}
\author[t]{A. Utrobicic}
\author[f,g]{I. Vai}
\author[e]{R. Veenhof}
\author[f,g]{P. Vitulo}
\author[o]{X. Wang}
\author[u]{S. White}
\author[m]{W. Xi}
\author[o]{Z. Zhang}
\author[o]{Y. Zhou}

\affiliation[a]{IRFU, CEA-Université Paris-Saclay, F-91191 Gif-sur-Yvette, France}
\affiliation[b]{Department of Physics, Aristotle University of Thessaloniki, GR-54124 Thessaloniki, Greece}
\affiliation[c]{CIRI-AUTH, GR-57001 Thessaloniki, Greece}
\affiliation[d]{Department for Medical Physics, Ludwig Maximilian University of Munich, 85748 Garching, Germany}
\affiliation[e]{CERN, 1211 Geneva 23, Switzerland}
\affiliation[f]{Dipartimento di Fisica, Università di Pavia, 27100 Pavia, Italy}
\affiliation[g]{INFN Sezione di Pavia, 27100 Pavia, Italy}
\affiliation[h]{Department of Physics and Astronomy, Stony Brook University, NY 11794-3800, USA}
\affiliation[i]{Institute of Nuclear and Particle Physics, NCSR Demokritos, GR-15341 Agia Paraskeui, Attiki, Greece}
\affiliation[j]{Helmholtz-Institut für Strahlen- und Kernphysik, University of Bonn, 53115 Bonn, Germany}
\affiliation[k]{Laboratório de Instrumentação e Física Experimental de Partículas (LIP), Lisbon, Portugal}
\affiliation[l]{Helsinki Institute of Physics, University of Helsinki, FI-00014 Helsinki, Finland}
\affiliation[m]{Jefferson Lab, Newport News, VA 23606, USA}
\affiliation[n]{Faculty of Electrical Engineering and Computing, University of Zagreb, 10000 Zagreb, Croatia}
\affiliation[o]{State Key Laboratory of Particle Detection and Electronics, University of Science and Technology of China, 230026 Hefei, China}
\affiliation[p]{Physikalisches Institut, University of Bonn, 53115 Bonn, Germany}
\affiliation[q]{Friedrich-Alexander-Universität Erlangen-Nürnberg, 91054 Erlangen, Germany}
\affiliation[r]{CEA-List, Diamond Sensors Laboratory, CEA-Saclay, F-91191 Gif-sur-Yvette, France}
\affiliation[s]{National Technical University of Athens, Athens, Greece}
\affiliation[t]{Ruđer Bošković Institute, 10000 Zagreb, Croatia}
\affiliation[u]{University of Virginia, Virginia, USA}

\emailAdd{alexandra.kallitsopoulou@cea.fr}

\abstract{We present a detailed performance study of a resistive PICOSEC Micromegas detector prototype, tested under highly energetic muon beam of 150\,GeV/c at the CERN SPS H4 beam line. This work provides a proof of concept for the use of resistive layer technology in gaseous timing detectors, demonstrating that robustness can be improved without compromising the excellent timing performance of PICOSEC Micromegas. Primary objective is to establish and validate a comprehensive analysis framework for timing and spatial performance evaluation, using a 10\,M$\Omega/\square$ resistive-plane prototype as the reference configuration. Using this framework
the 10\,M$\Omega/\square$ prototype achieved a timing resolution of $22.9 \pm 0.2$\,ps in single-pad events, and a core spatial resolution of $1.195 \pm 0.003$\,mm (X) and $1.197 \pm 0.003$\,mm (Y), while charge sharing across multiple pads enabled combined timing resolutions below 28\,ps. An additional prototype featuring a lower resistivity layer (200\,k$\Omega/\square$) was tested under identical conditions to assess the impact of resistivity on charge transport, timing, and spatial performance, reaching a timing resolution of $31.6 \pm 0.3$\,ps. Minor systematic spatial variations observed in the combined Signal Arrival Time (SAT) map are attributed to photocathode non-parallelism and readout PCB planarity imperfections, which are known to induce non-uniform drift fields affecting timing homogeneity. These studies benchmark the potential of resistive layers for gaseous timing detectors and provide a foundation for scalable designs with optimized timing and spatial resolution across diverse experimental applications.}

\keywords{Gaseous detectors, Timing detectors, Micromegas, Resistive Detectors, Spatial Resolution, Multipad-Analysis}

\begin{document}
\maketitle
\flushbottom

\section{Introduction}\label{sec:introduction}

The motivation for developing state-of-the-art detectors is to equip future experiments with high precision in time, energy, and position measurements. Among the many challenges faced by the field, precision timing stands out as a critical aspect of instrumentation technology. Accurate timing capabilities are essential for identifying the wide range of particles produced in environments characterized by high radiation flux and pile-up events. As modern facilities push toward increasingly intense particle beams and higher luminosities, the demand for detectors that can operate reliably under these conditions continues to grow \cite{Detector:2784893}.

The stringent requirements imposed by modern experiments push detector technologies to their operational limits. To meet the demands of high-luminosity environments, detectors must not only ensure excellent timing performance but also provide large-area coverage, resilience against aging effects, and the capability of multi-pad readout for precise tracking. In the domain of fast timing, solid-state detectors currently represent a benchmark technology \cite{Sadrozinski_2018}. However, promising alternatives are also emerging within the gaseous detector sector. Resistive Plate Chambers (RPCs), for instance, can achieve timing resolutions on the order of 30\,ps \cite{SANTONICO1981377, CERRONZEBALLOS1996132}, while members of the MicroPattern Gaseous Detector (MPGD) family typically reach a few nanoseconds \cite{GIOMATARIS}. Among them, Micromegas-based detectors stand out by combining competitive timing performance with additional advantages, such as scalability to large surfaces, cost-effectiveness, and high-rate capability.

The timing resolution of conventional Micromegas detectors is fundamentally limited by the diffusion of electrons within the large drift region and by the stochastic nature of ionization induced by the incident particle. These effects impose uncertainty on the conditions under which primary electrons are produced. The timing performance can be improved by suppressing the randomness of ionization and ensuring that all primary electrons are produced at the same distance from the mesh. 

\begin{figure}[hbt!]
  \centering
  \begin{subfigure}[b]{0.5\textwidth}
    \centering
    \includegraphics[width=\textwidth]{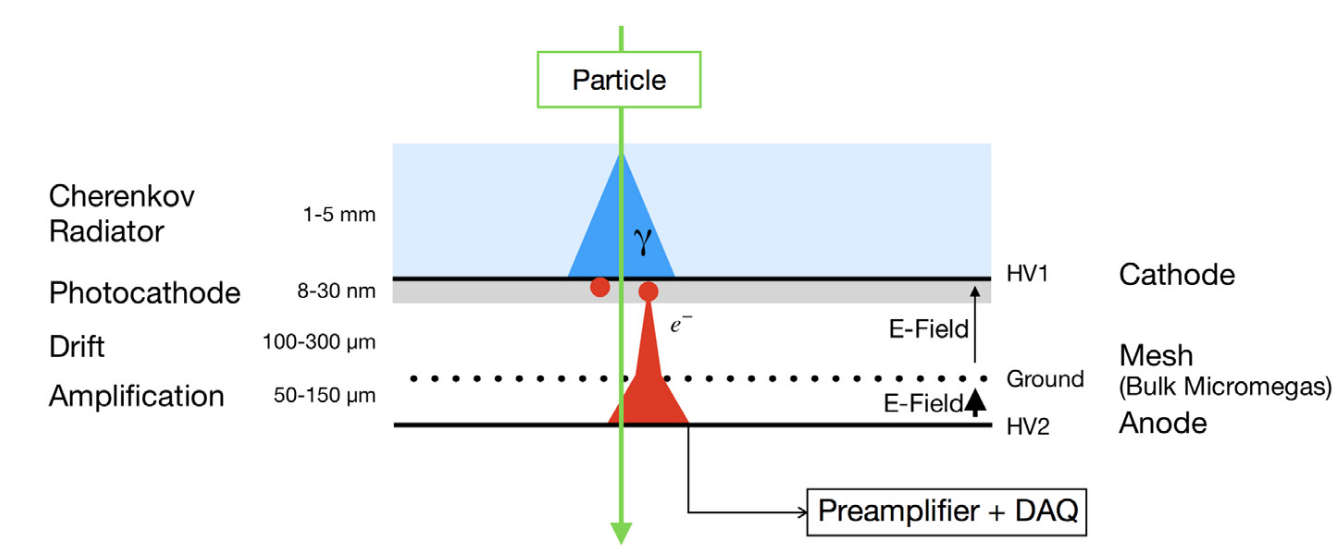}
    \caption{}
    \label{fig:picosec}
  \end{subfigure}
  \hfill
  \begin{subfigure}[b]{0.4\textwidth}
    \centering
    \includegraphics[width=\textwidth]{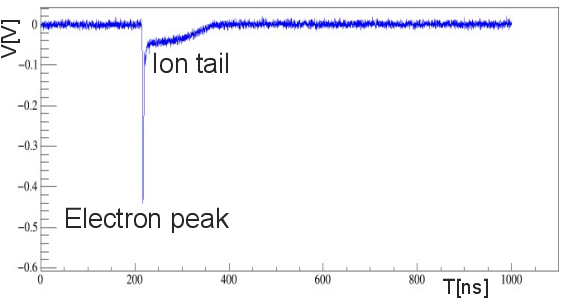}
    \caption{}
    \label{fig:picosec_waveform}
  \end{subfigure}
  \caption{(a) Graphical representation of a PICOSEC Micromegas detector \cite{BORTFELDT2018317}. (b) Typical PICOSEC Micromegas signal after the amplifier.}
\end{figure}

We introduce the PICOSEC Micromegas Detector concept to address these limitations, as described in Fig.\ref{fig:picosec}. This novel Micromegas-based detection system is designed to achieve charge-particle timing with a precision of tens of picoseconds. Unlike conventional Micromegas detectors, where ionization occurs randomly within the gas volume, the PICOSEC configuration employs a Cherenkov radiator. When a relativistic charged particle traverses the radiator, Cherenkov photons are emitted. A thin photo-converting material converts these photons into photoelectrons, ensuring that the photoelectrons enter the gas volume simultaneously and at a uniform distance from the mesh.

The PICOSEC Micromegas detector operates as follows: a relativistic particle crossing the 3 mm thick $\mathrm{MgF_2}$ Cherenkov radiator emits photons, as its velocity exceeds the speed of light in the medium. These Cherenkov photons strike the photocathode, releasing photoelectrons in a highly synchronized way. For 150\,GeV/c muons, the Cherenkov angle in MgF$_2$ is approximately 45°, producing a photon cone of approximately 6\,mm diameter at the photocathode. When entering a gas volume under a strong electric field, the photoelectrons immediately trigger charge multiplication, forming an avalanche. As the avalanche reaches the micromesh, part of the charge passes through into the amplification gap, where a higher electric field induces further multiplication. This process generates a measurable current on the anode, which is recorded by the readout electronics. By this mechanism, the timing jitter—typically a few nanoseconds in conventional Micromegas—is reduced to the tens-of-picoseconds level, enabling much higher precision. An example of a PICOSEC Micromegas Signal can be found in Fig.\ref{fig:picosec_waveform}. This signal is composed of the fast (700\,ps) electron peak and the slow (100\,ns) ion tail components.

The key distinction from standard Micromegas lies in the reduced drift (or conversion) gap, which has been shrunk from 3\,mm to 200\,\textmu m. This smaller gap both suppresses direct ionization by incident particles and increases the electric field in the drift region, which in turn raises the Townsend coefficient and enables the formation of early “preamplification avalanches.” These avalanches  stabilize the multiplication process before charges undergo final amplification in the gap below the mesh \cite{BORTFELDT2018317}.

The selection of an ideal gap is critical to ensuring the stability of preamplification avalanches, which directly correlates with improved timing resolution. A detailed study resulted in the fact that the avalanche formation drifts faster, with higher drift velocity, than its electron components, which means that the creation of the avalanche in the early stages of the drift gap will influence the timing resolution of the detector \cite{BORTFELDT2021165049}. These advantages are considered for the optimal operation of the PICOSEC Micromegas Detector. 

Since the first proof-of-concept, the PICOSEC Micromegas detector has demonstrated remarkable progress in timing performance. The initial prototype, was a single channel of 1\,cm diameter active zone operating in a gas mixture  $\mathrm{Ne}:\mathrm{CF_4}:\mathrm{C_2H_6}$ (80:10:10) at nominal temperature and pressure. It consisted of a 3\,mm $\mathrm{MgF_2}$ radiator window with 18\,nm CsI semitransparent protocathode on top of a 5.5\,nm Cr layer. This prototype achieved a record time resolution of 24\,ps with 150\,GeV/c muons, and a single-photoelectron resolution of 75 ps. Subsequent optimization, such as reducing the drift gap from 200\,\textmu m to 120\,\textmu m, improved the single-photoelectron resolution to 50\,ps. Advancements on the design of the single channel prototypes, focusing on operation stability, respecting the signal integrity, noise reduction, and reassembly features, lead to pushing even further the timing resolution to 13.3\,ps \cite{utrobicic2024singlechannelpicosecmicromegas}. 

Building on these results, the collaboration has focused on extending the technology to larger active areas while maintaining stability, robustness, and uniform timing. A 100-channel detector with a 100\,$\mathrm{cm^2}$ active area, reached time resolutions of 17 ps in the central pad region and 17–18 ps over most of the surface, while the biggest 400-channel prototype with 400\,$\mathrm{cm^2}$ active area reached 25\,ps with a uniformity of 33\% \cite{Utrobicic_2023, meng2025picosecmicromegasprecisetimingdetectors}. 

To develop more robust and modular prototypes that meet the demanding specifications of future applications, such as the ENUBET Project \cite{longhin2022enhancedneutrinobeamskaon} or the Muon Collider \cite{AIME2026170963}, this work pursues two main objectives. First, it establishes and validates a comprehensive analysis framework for timing and spatial performance evaluation, using a 10\,M$\Omega/\square$ resistive-plane prototype as the reference configuration. Second, it assesses the impact of resistivity on detector performance by comparing the reference prototype with a 200\,k$\Omega/\square$ variant, providing the first study of spatial resolution in resistive PICOSEC Micromegas detectors. A capacitive charge-sharing prototype \cite{Alviggi_2026, GNANVO2023167782} was also developed as part of the broader R\&D programme; its results will be reported in a dedicated future publication alongside simulation studies of the capacitive coupling and charge transport mechanisms \cite{kallitsopoulou:tel-05267379}. Together, these studies benchmark the potential of resistive layers for gaseous timing detectors and lay the foundation for scalable designs tailored to diverse applications. 

This paper is organized as follows. Section \ref{sec:resistive-developement} introduces the seven-pad resistive Micromegas prototype used for all measurements. Section \ref{sec:exp-set-up} describes the experimental setup and the standard waveform analysis procedure. Section \ref{sec:det-alignment} focuses on the detector alignment, while Section \ref{sec:timing-calibration} presents the timing performance of the individual pads. Section \ref{sec:combined-analysis} discusses the combined timing and spatial performance of the full detector assembly. Finally, Section \ref{sec:conclusion} summarizes the main conclusions.

\section{The 7-pad Resistive Prototype Development}\label{sec:resistive-developement}
\begin{figure}[hbt!]
 \centering 
 \includegraphics[width=0.5\textwidth]{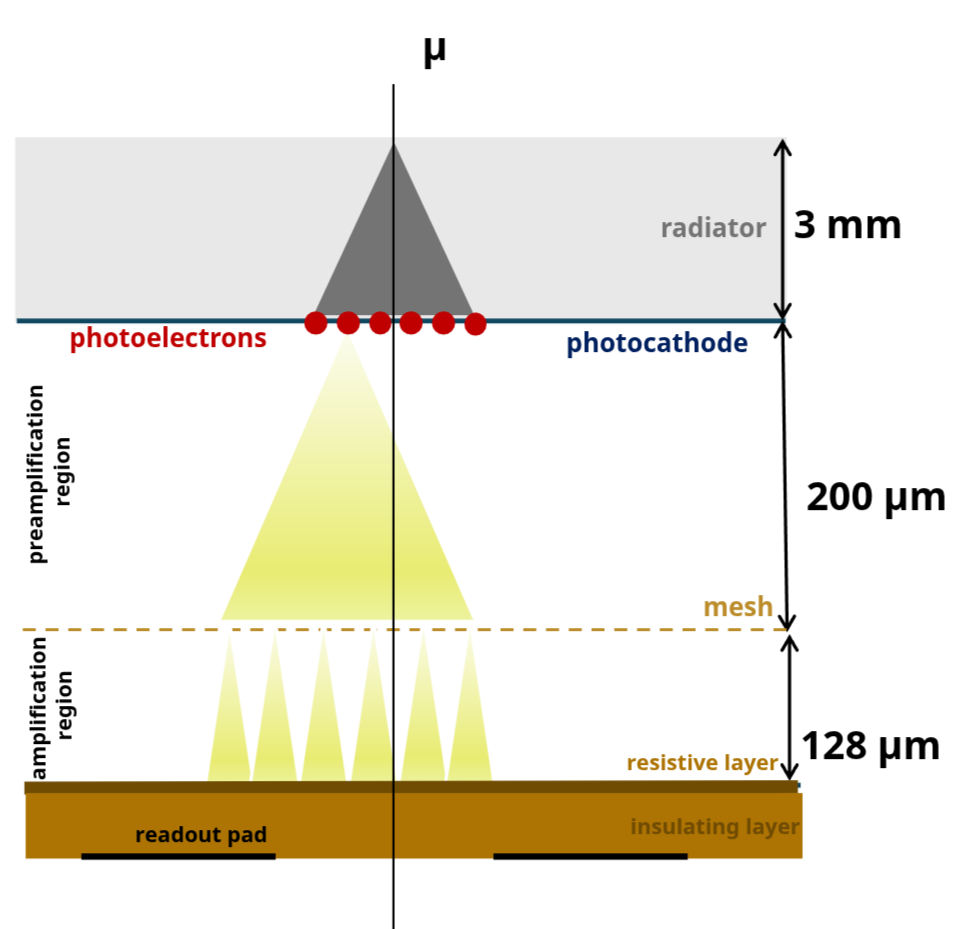}
\caption{Schematic description of the resistive PICOSEC Micromegas detector layout.}\label{fig:resistive_schema}
\end{figure}

To mitigate the destructive impact of discharges in the PICOSEC Micromegas detector, a resistive anode structure was implemented in the detector configuration. This approach efficiently quenches discharges, suppressing the streamer–spark transition while preserving high gain and stable operation under demanding conditions. The design, illustrated in Fig.~\ref{fig:resistive_schema}, incorporates a thin Diamond-Like Carbon (DLC) resistive layer, electrically insulated from the pad electrodes by a 50\,\textmu m polyimide layer and a 10\,\textmu m adhesive layer.

All prototypes tested in this study were based on the same readout geometry, consisting of seven hexagonal pads with an outer diameter of 10\,mm, an inner diameter of 8.66\,mm, and an inter-pad spacing of about 200\,\textmu m to reduce capacitive coupling and crosstalk. The resistive-layer contact and the mesh termination were routed outside the active area, to further reduce the probability of discharges in the sensitive region.

Within this common geometry, different resistive-layer implementations were investigated. Two surface resistivities were tested for the DLC films, 10\,M\(\Omega\)/\(\square\) and 200\,k\(\Omega\)/\(\square\), chosen to represent extreme values in the relevant range. The configuration employed a uniform DLC foil, in which a continuous resistive layer covered all pads, enabling charge dispersion and providing intrinsic protection against discharges.

\begin{figure}[hbt!]
\centering 
\includegraphics[width=0.8\textwidth]{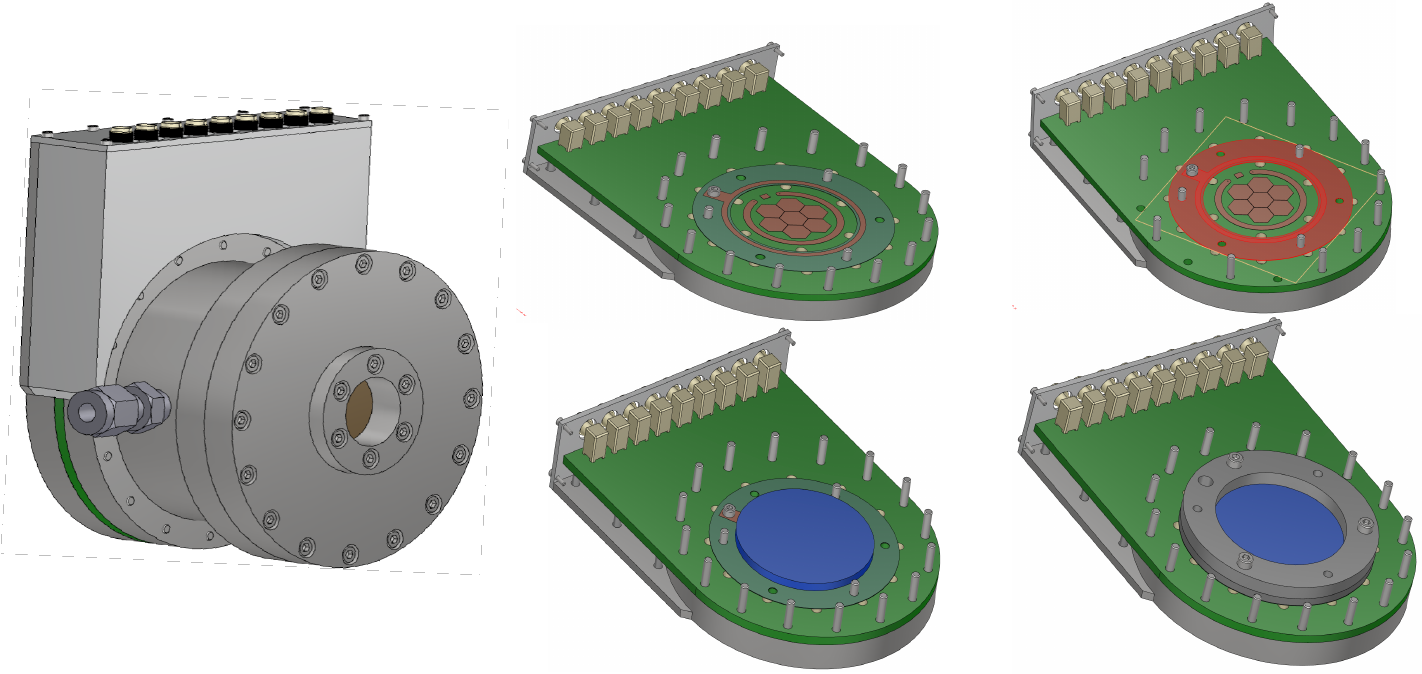}
\caption{Left: Full detector assembly on the chamber. Right (from top to bottom): Step-by-step assembly procedure, from the readout PCB, the drift copper ring, which preserves electrical contact and defines the drift gap size, the 5\,cm  \diameter  crystal placement on top of the drift spacer, and the PEEK support.}
\label{fig:7-pad-assembly}
\end{figure}

All prototypes were integrated into a common detector housing, illustrated in Fig.~\ref{fig:7-pad-assembly}. The assembly procedure placed a copper drift spacer on the readout PCB, secured with screws to provide both electrical connection (via PCB traces and an SMA connector) and precise definition of the drift gap. A 5\,cm diameter, 3\,mm-thick MgF\(_2\) radiator, acting as the Cherenkov radiator and photocathode, was mounted on the spacer and held in place by a PEEK ring to ensure mechanical stability.  

For comparison reasons, the prototypes were tested using an identical drift gap of 150\,\textmu m to ensure compatibility. For assessing optimal timing performance, a standardized photocathode reference was employed, comprising an 18\,nm CsI layer deposited on a 3.3\,nm chromium (Cr) conductive substrate. A complete table with all the information about the prototypes developed and tested is given in Tab. \ref{tab:7-pad-res-prototypes}. 

\begin{table}[hbt!]
\centering
\caption{Summary of the tested configurations, showing key design parameters and common operating conditions.}
\label{tab:7-pad-res-prototypes}
\small
\setlength{\tabcolsep}{4pt}
\renewcommand{\arraystretch}{1.15}

\begin{tabularx}{\linewidth}{|
>{\centering\arraybackslash}p{0.10\linewidth}|
C|
>{\centering\arraybackslash}p{0.13\linewidth}|
>{\centering\arraybackslash}p{0.25\linewidth}|
>{\centering\arraybackslash}p{0.20\linewidth}|}
\hline
\textbf{Resistivity} &
\textbf{Architecture / Layout} &
\textbf{Drift/Ampl. gap[\textmu m]} &
\textbf{Photocathode} &
\textbf{Field configuration Drift / Ampl. [kV/cm]} \\ \hline

10\,M$\Omega$/\(\square\) &
Plane DLC (resistive foil) on top of the hexagonal pads &
150/128 &
18\,nm CsI on 3.3\,nm Cr &
36.6/21.4 \\ \hline

200\,k$\Omega$/\(\square\) &
Plane DLC (resistive foil) on top of the hexagonal pads &
150/128 &
18\,nm CsI on 3.3\,nm Cr &
36.6/21.4 \\ \hline

\end{tabularx}
\end{table}

\section{Experimental Setup and Standard Waveform Analysis}\label{sec:exp-set-up}
The prototypes were tested during the 2023 RD51/DRD1 collaboration beam test campaigns at CERN. The measurements took place at the H4 beamline of the SPS using 150\,GeV muons.

\begin{figure}[hbt!]
 \centering 
 \includegraphics[width=0.8\textwidth]{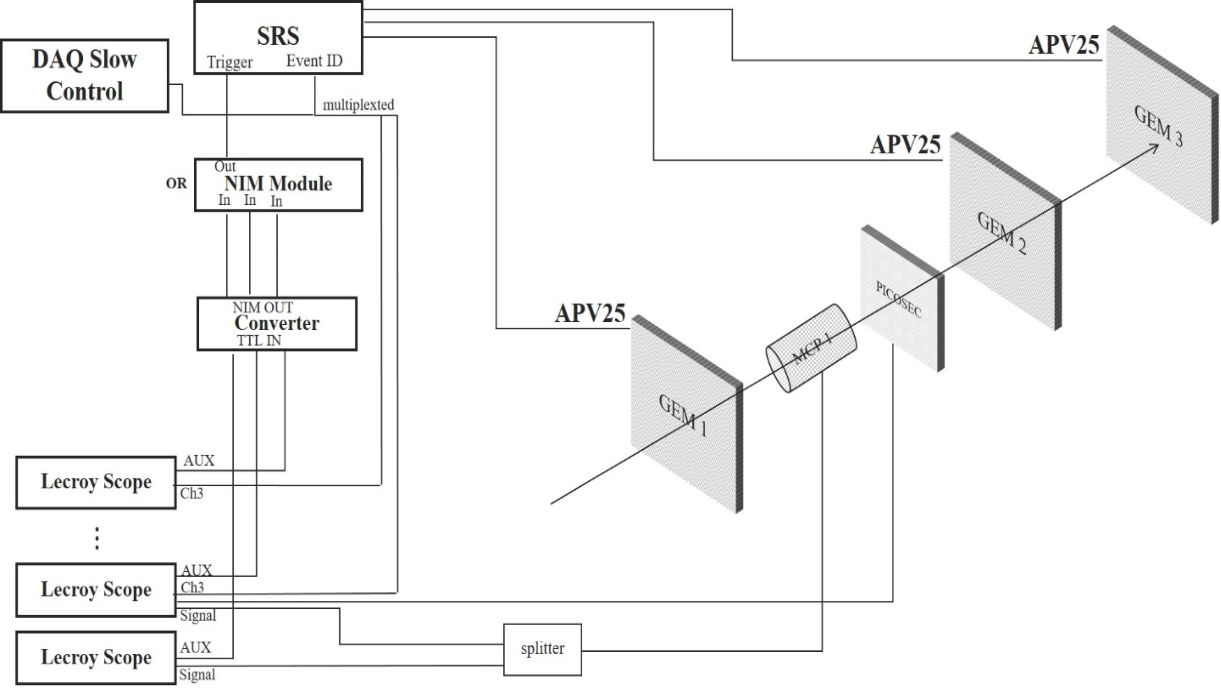}
\caption{Block diagram and sketch of the electronic modules used to provide NIM signal to trigger the DAQ.}\label{fig:triple_gem}
\end{figure}

The experimental setup consisted of a beam telescope capable of accommodating up to six PICOSEC-Micromegas prototypes, where several prototypes from the PICOSEC-Micromegas collaboration were installed simultaneously during the beam test campaigns. The measurements reported in this paper are those recorded by the 7-pad resistive prototypes described in Section~\ref{sec:resistive-developement}. The telescope included a tracking system made of three triple-GEM detectors and an MCP-PMT\footnote{Hamamatsu Microchannel Plate Photomultiplier Tube (MCP-PMT) R3809U-50, \url{https://www.hamamatsu.com/jp/en/product/optical-sensors/pmt/pmt_tube-alone/mcp-pmt/R3809U-50.html}} serving as precise timing references. Fig.\,\ref{fig:triple_gem} illustrates the trigger and readout scheme of the setup, showing a single PICOSEC detector as representative of the connection scheme used for each prototype. The GEMs \footnote{Gas mixture used was 70\% Ar and 30\% $\mathrm{CO_2}$ at NTP.}, provided two-dimensional hit reconstruction, with signals digitized by the SRS system \cite{martoiu2013development}. Particle tracks were reconstructed offline \cite{bortfeldt2014development} using cluster analysis and linear fits, assuming straight trajectories due to the high beam momentum and absence of magnetic fields.

Trigger signals were derived from the MCP-PMT and distributed to both the digitizers and the trigger logic of the SRS. As shown in Fig.\,\ref{fig:triple_gem}, the MCP-PMT signal was split and distributed to all oscilloscopes used for waveform acquisition, while the AUX outputs were discriminated and converted to a NIM logic pulse. The replicated MCP logic signals were combined in a NIM OR module to form the global trigger. This trigger initiated the SRS readout, which in turn generated a unique event identifier redistributed to all oscilloscopes to ensure event-by-event synchronization between the tracker data and the digitized PICOSEC-Micromegas waveforms.
The APV25 data \cite{martoiu2013development} were acquired with the DATE software \footnote{CERN. ALICE DAQ and ECS Manual (2010), \url{https://cds.cern.ch/record/689027/files/INT-1998-04.pdf}}, while the waveforms from the prototypes were recorded with fast oscilloscopes. In offline analysis, the signals from the prototypes were matched to reconstructed particle trajectories, allowing for precise spatial and timing performance studies.

The signals from the seven PICOSEC-Micromegas pads pass through custom-made seven-channel board-structured preamplifiers. These amplifier cards are being used with a gain of 38\,dB, a bandwidth of 650\,MHz, and built-in discharge protection based on the RF amplifier design of \cite{Hoarau_2021}. PICOSEC-Micromegas signals are sampled with oscilloscopes from Lecroy Waverunner Series\footnote{Tedelyne LeCroy. Waverunner 8000 Series Datasheet (2018)}, operating in 10\,GS/s sampling rate. 

The sampled waveforms, along with reference and trigger signals, were stored as binary files by the oscilloscope and processed using a dedicated analysis procedure. 

The offline analysis followed a standardized workflow to ensure consistent signal reconstruction and timing extraction across all prototypes. The digitized waveforms from the PICOSEC-Micromegas detectors and the MCP-PMT reference were first pre-processed to correct for baseline fluctuations and suppress high-frequency noise. The baseline level was estimated from the pre-trigger region of each waveform and subtracted on an event-by-event basis. A three-point smoothing algorithm was then applied to reduce residual oscillations without distorting the pulse shape. This procedure ensures a robust and independent definition of the pulse start and end points (pulse boundaries), which is essential for extracting key parameters, such as the integrated charge \cite{kallitsopoulou:tel-05267379}.

The signal arrival time (SAT) was determined using a Constant Fraction Discrimination (CFD) technique based on a logistic fit to the leading edge of each pulse. The timing was extracted at 20\% of the fitted amplitude, providing sub-sample precision and minimizing amplitude-dependent timing bias.  The same method was applied to both the reference MCP-PMT and the detector under test. The SAT distribution was obtained by subtracting the PICOSEC-Micromegas timing from that of the MCP-PMT reference, both digitized by the same oscilloscope \cite{BORTFELDT2018317}.

Although this approach is designed to avoid systematic effects, the observed correlation between the SAT and the signal amplitude arises from intrinsic detector physics. Specifically, the microscopic development of the avalanche introduces a dependence of the signal formation time on the initial photoelectron drift. As individual electrons drift slower than the propagating avalanche front, the time required to initiate the multiplication process manifests as an apparent time walk. This effect is unrelated to the analysis method itself but reflects the stochastic nature of avalanche formation. The dependence of SAT on signal amplitude was therefore parameterized and corrected offline. The corrected SAT distribution was then fitted with a Gaussian function, and the standard deviation of this fit represents the intrinsic timing resolution of the detector \cite{BORTFELDT2021165049}.

For a more accurate parameterization of this behavior, two observables can be considered: the electron-peak charge and the total charge of the waveform. The electron-peak charge is defined as the integral of a fitted function representing the fast component of the signal. Specifically, the electron peak, which begins at the triggering point and extends up to approximately 6\,ns after the maximum amplitude, is fitted using a double-sigmoid function:

\begin{equation}\label{eq:double_logistic}
    f(x;p_0,p_1,p_2,p_3, p_4, p_5) = p_3 + \frac{p_0}{1+e^{-(x-p_1)p_2}} \times \frac{1}{1+e^{-(x-p_4)p_5}}
\end{equation}
where $p_0$ and $p_3$ are the maximum and minimum values of the amplitude, $p_1$ is the half point of the rise-time , $p_2$ is the steepness of the fitted function , and similarly, $p_4$ and $p_5$ are the half point and the steepness of the falling edge, respectively.

However, the total integrated charge was found to be a more robust and reliable observable, since determining the exact endpoint of the electron peak is often challenging. In particular, accurately identifying the transition between the electron peak and the onset of the ion tail can be difficult due to signal fluctuations. Pile-up candidates are identified independently by monitoring the derivative of the 5\,ns moving integral: pulses for which the derivative exceeds 20\% of the maximum derivative value are flagged as pile-up and excluded from the subsequent analysis. The endpoint of each clean, single-pulse waveform was determined using the cumulative sum of the waveform. Within a target window of approximately 150\,ns from the triggering point — corresponding to twice the typical pulse duration — the cumulative sum is computed over the baseline-subtracted waveform. The local minimum of the cumulative sum within this window corresponds to the point where the pulse tail begins to flatten and the integrated signal stops growing. This minimum is used as the upper integration bound for total charge calculation \cite{kallitsopoulou:tel-05267379}.

Measuring the timing performance of large-area detectors poses a challenge when selecting a reference device, which must provide both superior resolution and spatial uniformity compared to the detector under test. Although large-area MCP-PMTs are available, their Signal Arrival Time uniformity must be well understood to avoid introducing biases. Previous studies \cite{Bortfeldt_2020} showed that the MCP-PMT time resolution is below 6\,ps at the center but deteriorates rapidly toward the edges. To mitigate this effect, only the central region of the MCP was used in the present measurements.  

For those measurements, the MCP was mounted on a movable stage and scanned over the detector area in a 20$\times$20\,mm$^2$ grid with 2.5\,mm steps, ensuring about 75\% overlap, covering an area around the center of the central pad of the detector. By setting a sufficiently high threshold, to the MCP-PMT signal, events from the outer regions were excluded, and uniform data collection was achieved through synchronization with the number of recorded events. The data used in this work, are collected during long runs, where more than 500,000 triggers were collected with the detectors operated under stable operation, with the field configurations described in Tab. \ref{tab:7-pad-res-prototypes}.

\section{Detector Alignment}\label{sec:det-alignment}
To accurately study the timing performance of the multi-pad PICOSEC-Micromegas prototypes, as mentioned in \cite{AUNE2021165076}, it is essential to determine the precise position of each readout pad with respect to the global tracking reference. This alignment ensures that charge sharing and timing correlations between neighboring pads are correctly interpreted in the beam reference frame. The same method was used for all the different prototypes tested, but for simplicity, we describe in detail using the data recorded by the central pad of the 10\,M\(\Omega\)/\(\square\) prototype. 

The procedure begins with the alignment of each detector and pad to the tracking system, using the trajectories reconstructed by the GEM telescope. For every pad, a two-dimensional map of the average collected charge on that pad is constructed by correlating the reconstructed track impact positions with the measured charge on that pad. Since the beam uniformly illuminates the active area, the charge-weighted mean positions provide an unbiased first estimate of each pad center. 

\begin{figure}[hbt!]
\centering 
\includegraphics[width=0.8\textwidth]{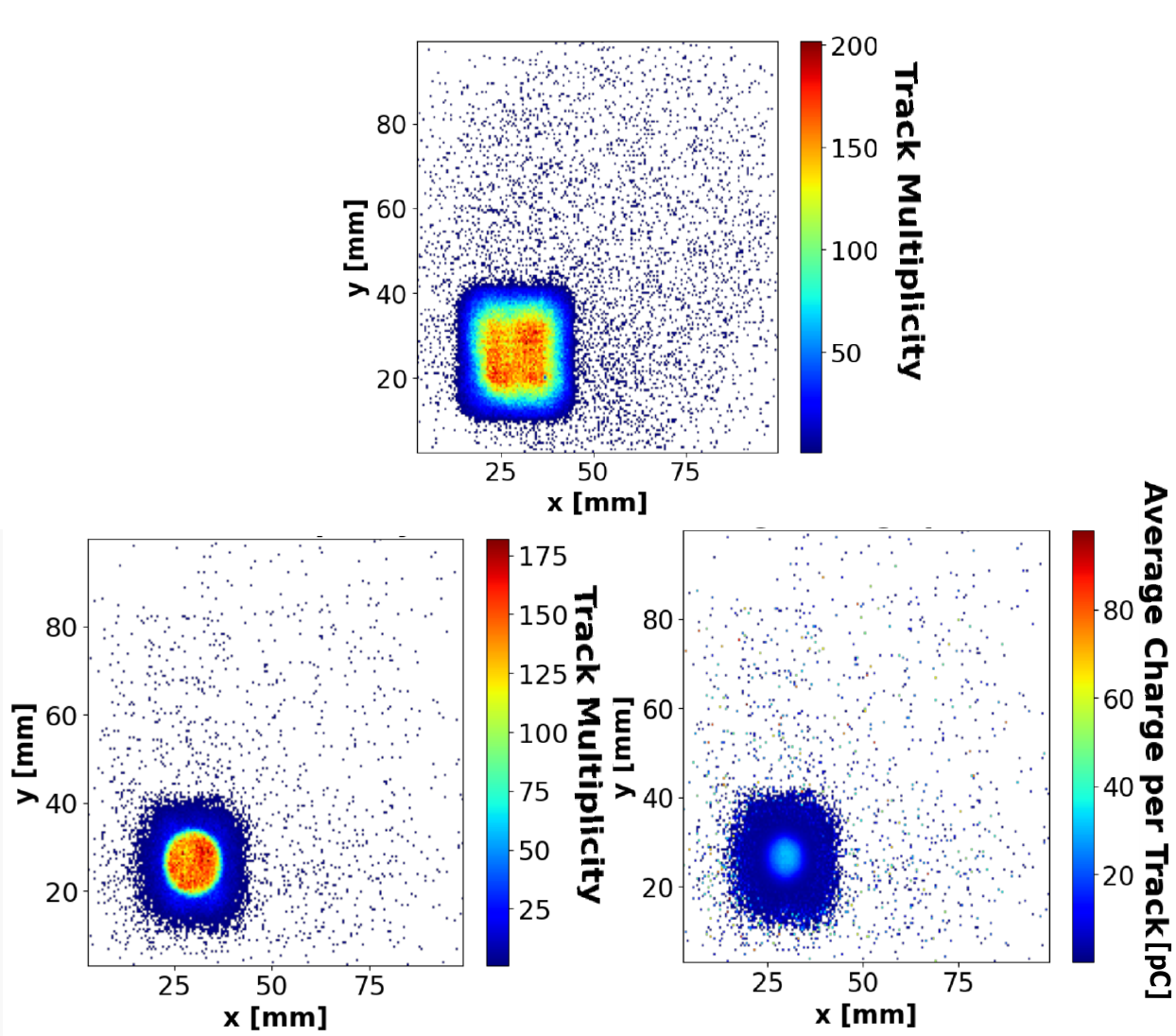}
\caption{Top: 2D spatial distribution of reconstructed tracks as measured by the MCP-PMT reference detector during the scan; the spatial extent reflects the MCP-PMT scan geometry, not a fiducial selection on the detector. Bottom, for pad-0: (Left) number of reconstructed tracks per bin ($N_i$); (Right) average charge per track (\added{$\bar{Q}_{xy} = \sum_{j=1}^{N_i} q_j / N_i$}), where the charge-weighted distribution highlights the position of pad-0.}
\label{fig:track-distributions}
\end{figure}

A refined determination of the pad centers is achieved through a multistage algorithm that combines binning, charge filtering, and charge-weighted fitting. The active area of the pad plane is divided into uniform bins of 0.5\,mm $\times$ 0.5\,mm, and for each bin the number of reconstructed tracks and the the total collected charge on that pad under study is computed. The top panel of Fig. \ref{fig:track-distributions} shows the spatial distribution of the beam as measured by the MCP-PMT reference detector during the scan. The boundary of the distribution reflects the geometrical acceptance and scan range of the MCP-PMT and does not correspond to a fiducial selection on the detector geometry. For pad-0, the bottom-left panel of the same figure shows the number of reconstructed tracks per bin associated with signals on the pad under study ($N_i$), while the bottom-right panel shows the corresponding average charge per track, the bin-averaged charge:
\[
\bar{Q}_{xy} = \added{\frac{\sum_{j=1}^{N_i} q_j}{N_i}}
\]
where \added{$q_j$ is the charge recorded on pad-0 for track $j$ within bin $i$,} and $N_i$ the number of tracks in the bin. This quantity is used for the alignment, and its spatial distribution highlights the pad-0 position within the detector plane.

To suppress the impact of statistical noise, only bins fulfilling basic quality conditions are retained: a minimum number of tracks ($T_\mathrm{min}=20$ by default) and a minimum average charge ($Q_\mathrm{min}=6$\,pC for these datasets). The value of $Q_\mathrm{min}$ is dataset-dependent and is chosen to be as low as possible while still ensuring that the resulting charge profiles are sufficiently symmetric around the pad centre to be reliably fitted with a symmetric polynomial. The default code value of 4\,pC is a conservative lower bound; for the datasets used here, 6\,pC was found to optimally balance noise rejection and signal acceptance. This filtering isolates the physically relevant signal region while excluding bins dominated by noise or edge effects.

\begin{figure}[hbt!]
\centering 
\includegraphics[width=0.9\textwidth]{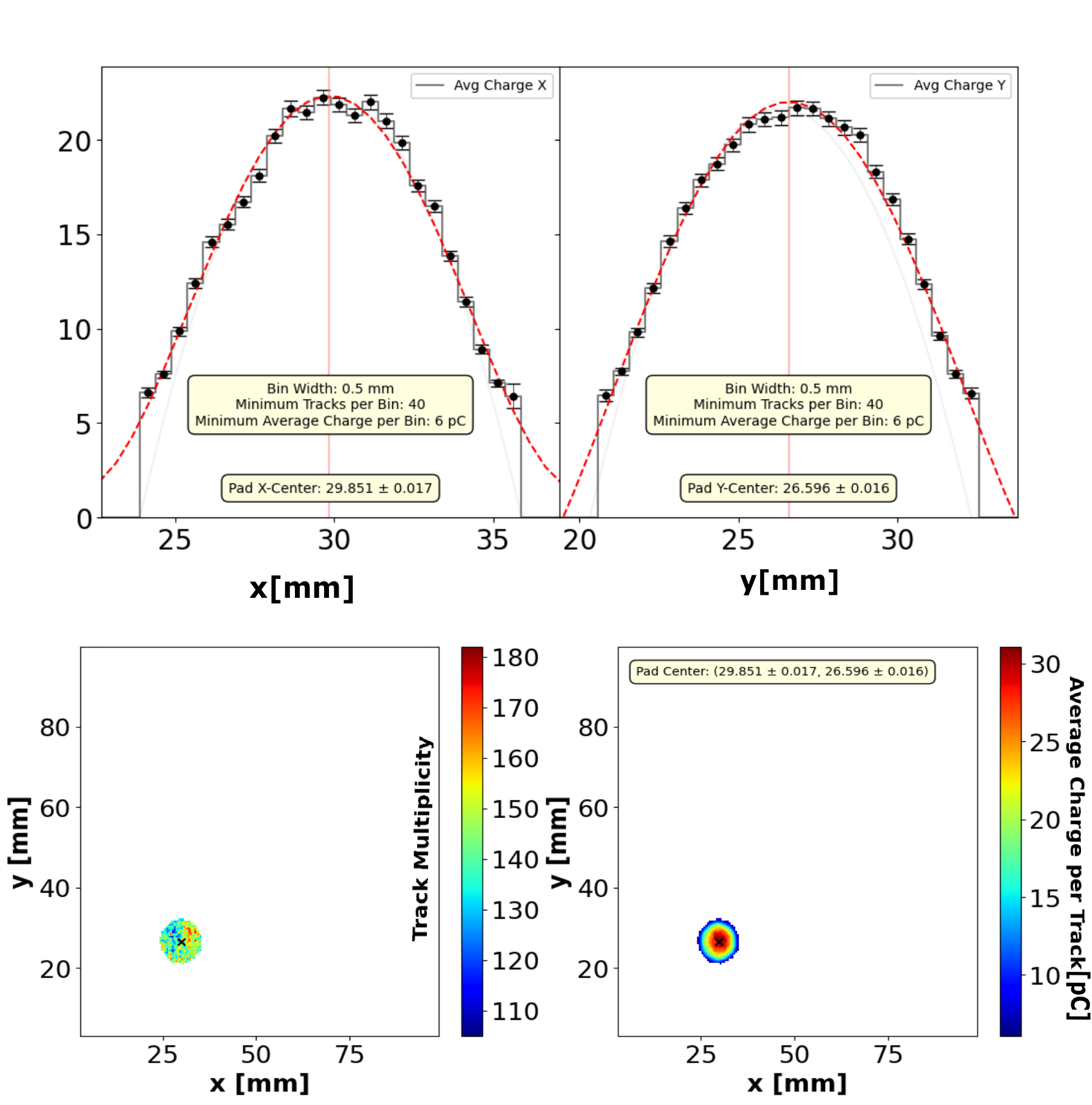}
\caption{Top: X and Y projections of the average charge distribution of pad-0. The dashed red line corresponds to a second-order even polynomial that fits the symmetric distribution, and gives the $x_c, y_c$ information. Bottom: The filtered 2D distributions for tracks (left) and average charge (right) on the pad-0 position. With x the  $x_c, y_c$ position of the center of the pad as estimated by the alignment method.}
\label{fig:10-MO-pad0-center-projections}
\end{figure}

The filtered two-dimensional charge distribution is then projected along the $x$ and $y$ axes to obtain one-dimensional charge profiles, $\bar{Q}_x(x_i)$ and $\bar{Q}_y(y_j)$. These distributions are expected to be approximately symmetric around the pad center, as shown in the top panel of Fig.\ref{fig:10-MO-pad0-center-projections}. To determine the precise centroid, each profile is fitted with a symmetric second-order polynomial of the form
\[
f(x) = a(x - x_0)^2 + b,
\]
where $x_0$ (and analogously $y_0$) corresponds to the position of the maximum charge density, this value represents the charge-weighted centroid of the pad response function, which we interpret as the pad’s geometric center in the beam reference frame.  

Each fit is weighted by the statistical uncertainty of the charge in each bin, $\sigma_{\bar{Q}} = \bar{Q}/\sqrt{N_\mathrm{tracks}}$, ensuring that bins with higher statistics dominate the fit. The covariance matrix from the polynomial fit provides the uncertainty on the extracted center coordinates, typically at the level of a few tens of micrometers.

The bottom panel of Fig. \ref{fig:10-MO-pad0-center-projections}
shows the exact same quantities as the bottom panel of Fig. \ref{fig:track-distributions} for pad-0, after applying the track and charge-quality selections described above, which suppress edge-sharing and low-statistics bins.

This procedure yields a robust and data-driven estimate of the pad centers and their relative alignment within the detector plane. The final coordinates, expressed in the global beam frame, are used as reference positions for subsequent timing and spatial correlation analyses across pads.

Accurate spatial reconstruction of charged-particle hits within the detector plane requires that the geometrical model of the detector be aligned with the measured positions of its readout pads. Small mechanical tolerances, mounting deviations, or rotations between the physical detector and the global beam frame can lead to systematic offsets if not properly corrected. This section details the alignment and angular calibration procedure, which enables the transformation from the experimental pad coordinates to a normalized local detector frame suitable for charge-sharing and timing analyses.

The detector is modeled as a regular array of hexagonal pads, each representing an independent readout element. This geometry is implemented computationally in a custom Python ( \texttt{HexDetector}) class, developed for the analysis, which constructs the hexagonal lattice by recursively placing neighboring pads at fixed angular offsets of 60\textdegree. Each pad is characterized by its inner radius $r_\mathrm{inner} = 4.33$\,mm, defining the distance from the pad center to the midpoint of one edge. The corresponding outer radius is given by $r_\mathrm{outer} = \frac{2}{\sqrt{3}}r_\mathrm{inner} = 5.0$\,mm, consistent with the 10\,mm outer diameter.

In this regular hexagonal lattice, all six nearest-neighbour pad centres are equidistant. The centre-to-centre distance in the model is $2r_\mathrm{inner} = 8.66$\,mm, corresponding to adjacent pads sharing a common edge. Accounting for the physical inter-pad spacing of $\sim$200\,\textmu m, the centre-to-centre distance in the physical detector is approximately 8.86\,mm.

A 7-pad configuration is initialized by placing the central pad at the origin $(0,0)$ in local coordinates, and its six neighbors at the prescribed hexagonal offsets. The measured pad centers from calibration are then imported as $(x_i^{meas}, y_i^{meas})$, forming the experimental reference for the alignment.

To determine the global translation and rotation between the physical detector and the idealized model, a non-linear least-squares minimization is performed. The objective function is defined as:

\begin{equation}
    \chi^2 (x,y,\theta) = \sum_{i=1}^7 \left[\left(\frac{x_i^{meas} - x_i^{det}(x,y,\theta)}{\sigma_{x_i}}\right)^2 + \left(\frac{y_i^{meas} - y_i^{det}(x,y,\theta)}{\sigma_{y_i}}\right)^2 \right],
\end{equation}

where $(x_i^{det}(x,y,\theta), y_i^{det}(x,y,\theta))$ are the detector model predictions after applying a translation $(x,y)$ and rotation $\theta$. The initial guesses for $(x,y,\theta)$ are derived from approximate detector placement measurements. The minimization is carried out using the \texttt{scipy.optimize.minimize} function, from the SciPy specific computing library \cite{virtanen2020scipy} with bounded parameters to ensure physical realism ($|x|, |y| < 15$\,mm, $|\theta| < 0.2$\,rad).  

\begin{figure}[hbt!]
\centering 
\includegraphics[width=0.8\textwidth]{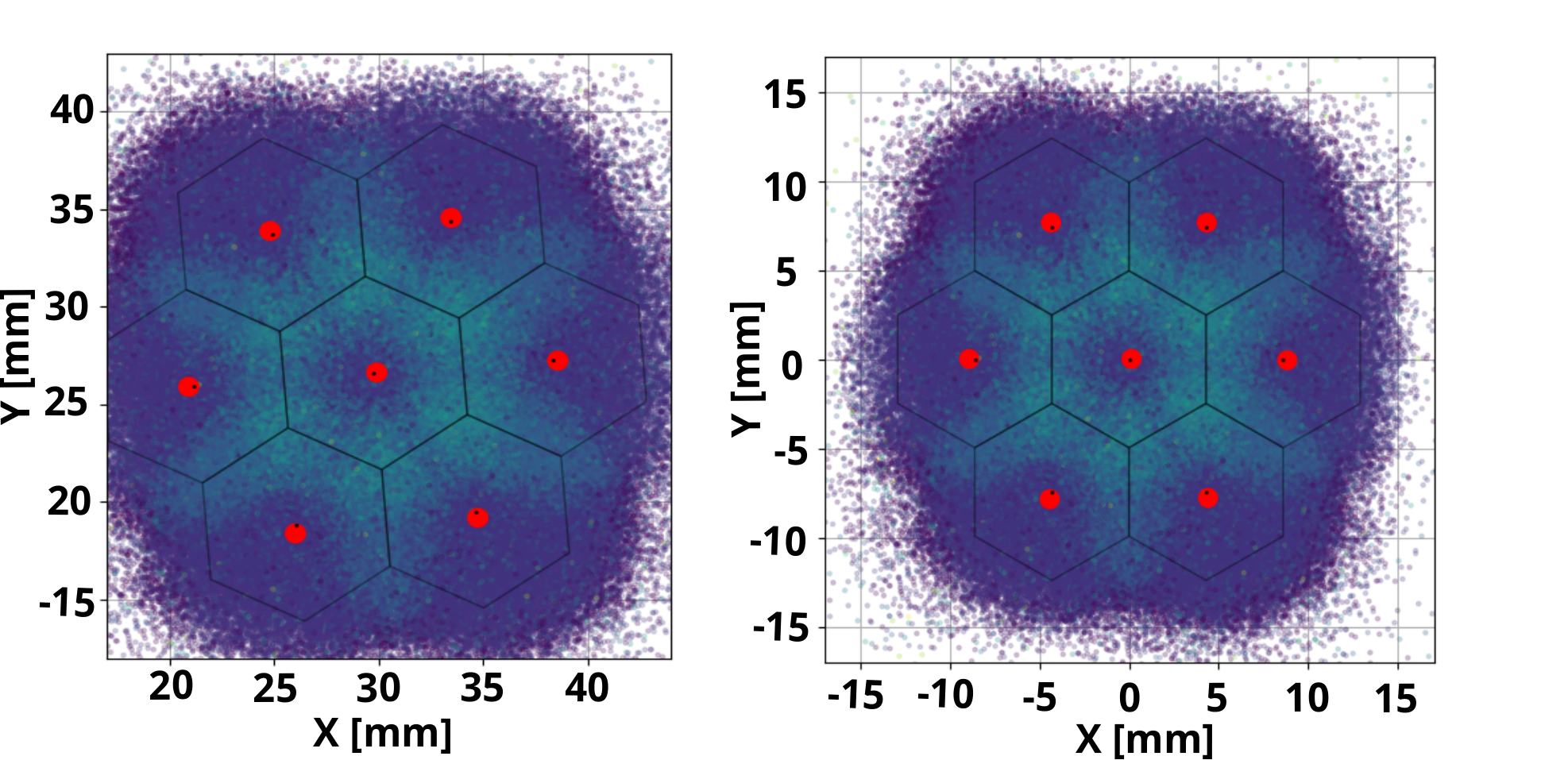}
\caption{(Left) Track hit positions in the global beam frame. (Right) Track hit positions after applying the full rigid-body alignment — translation by $(x^*, y^*)$ and rotation by $\theta^* = 5.5^\circ$ — into the local detector frame. Red markers indicate the reconstructed pad centres derived from charge-weighted hit positions; black dots show the expected geometrical pad centres from the idealized model. Marker sizes carry no physical meaning.}
\label{fig:rotating-tracks}
\end{figure}

The best-fit parameters $(x^*, y^*, \theta^*)$ correspond to the global transformation that minimizes the residuals between measured and modeled pad centers. In the case of the 10\,M\(\Omega\)/\(\square\), as shown in Fig.\ref{fig:rotating-tracks}, the best-fit values are $x^* = 29.78$\,mm, $y^* = 26.57$\,mm, and $\theta^* = 5.5^\circ$. This correction compensates for the slight tilt of the detector relative to the beam reference frame. The per-pad residuals after alignment range from 82\,\textmu m to 369\,\textmu m (mean 264\,\textmu m), confirming the validity of the global rigid-body approximation at sub-millimeter level.

\subsection*{Application of the Angular Correction}

After obtaining the optimal transformation parameters, all measured track and hit positions are mapped into the detector’s local coordinate system. Each point $(x, y)$ is translated by the best-fit detector centre $(x^*, y^*)$ and then rotated by $\theta^*$, expressed as the combined rigid-body transformation:
\begin{equation}
    {\color{red}
    \begin{pmatrix} x_{\mathrm{rot}} \\ y_{\mathrm{rot}} \end{pmatrix}
    =
    \begin{pmatrix} \cos\theta^* & -\sin\theta^* \\ \sin\theta^* & \cos\theta^* \end{pmatrix}
    \begin{pmatrix} x - x^* \\ y - y^* \end{pmatrix}.
    }
\end{equation}
The resulting coordinates $(x_{\mathrm{rot}}, y_{\mathrm{rot}})$ define a normalised local coordinate system in which the pad centres coincide with the idealised hexagonal geometry, ensuring that the subsequent timing and signal reconstruction analyses are free from geometric distortions caused by angular misalignment.

All steps of the procedure are fully reproducible using the provided \texttt{HexDetector} class and the corresponding alignment script in Python. The method ensures that the geometrical model of the detector matches the measured pad layout with a mean per-pad residual below 300\,\textmu m, sufficient for the subsequent timing and signal reconstruction analyses.

\section{Timing Characteristics of Individual Pads}\label{sec:timing-calibration}

To ensure that the derived timing properties of each pad reflect the intrinsic response of the detector and not artifacts of the data acquisition or geometry, several quality cuts were introduced. The raw pulse amplitudes were first inspected to identify saturation effects arising from the dynamic range limit of the oscilloscope digitizer. This selection removes high-amplitude events where the signal was cut by the digitizer, improving the consistency of the CFD time extraction.

The reconstructed track parameters, provided by the external triple-GEM telescope, were used to ensure that only well-defined particle trajectories were associated with valid PICOSEC events. The $\chi^2$ value from the telescope track fit was used as a measure of the track quality. Events with $\chi^2 > 3$ were rejected, corresponding to about 7\% of the total statistics. This cut eliminates poorly reconstructed tracks that could lead to mismatched hit positions and hence deterioration of the pad center estimation.

To minimize the contribution of induced signals on neighboring pads, through the resistive layer, only events reconstructed within the active radius of each pad were considered. Based on the hexagonal pad geometry, the fiducial area was defined as a circular region of radius $r < 5$\,mm around the pad center. This selection ensures that the waveform corresponds to a direct local response rather than a resistive or capacitive coupling from adjacent electrodes.

Following the timing analysis procedure described previously, based on the Constant Fraction Discrimination (CFD) method, we extracted the SAT for each pad and investigated its dependence on the total charge. Several parameterization approaches were tested to describe the mean SAT as a function of charge, including Gaussian fitting and median-based extraction per charge bin \cite{kallitsopoulou:tel-05267379}. Both methods yielded consistent results, and the Gaussian mean was adopted as the standard estimator.

The dependence of the SAT on total charge, $\mathrm{SAT}(q)$, was modeled using a double-exponential function plus a constant term, as expressed by:

\begin{equation}\label{eq:double_expo}
f(q; p_0, p_1, p_2, p_3, p_4) = e^{(p_0 q + p_1)} + e^{(p_2 q + p_3)} + p_4
\end{equation}

\begin{figure}[hbt!]
\centering 
\includegraphics[width=0.9\textwidth]{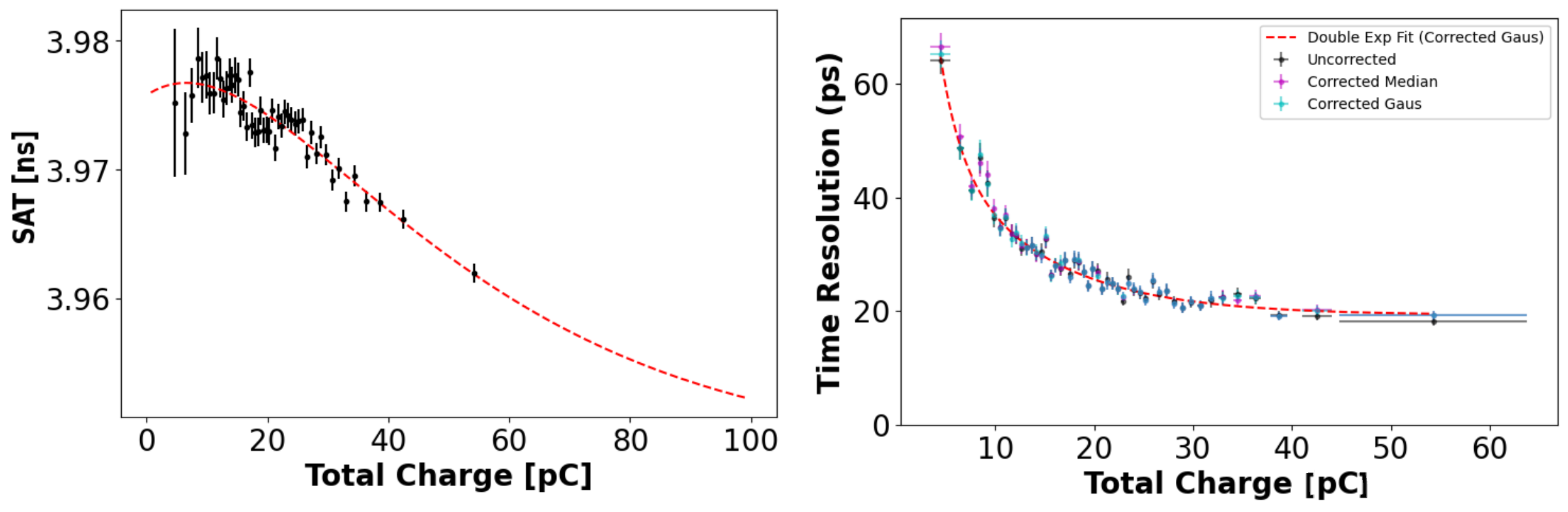}
\caption{(Left) Dependence of SAT versus total charge, after the applied cuts for saturation, $\chi^2$, and fiducial cut at radial distance of 5\,mm and with their double exponential fit function distribution. (Right) Timing Resolution as a function of Total Charge for the pad-0 region of the 7-pad detector. With black, the raw values, and with magenta and cyan, the corrected values based on their SAT correction.}
\label{fig:timing-resolution-pad0}
\end{figure}

This parameterization captures the characteristic trend of the time-walk effect across the charge spectrum. The systematic dependence of SAT on signal amplitude was then corrected by subtracting the fitted function $f(q)$ from the measured data points on an event-by-event basis, yielding the corrected SAT distributions. Using the same charge-binning procedure, the timing resolution was extracted as the gaussian $\sigma$ of the SAT distributions in each charge bin, both before and after time-walk correction, as presented in Fig. \ref{fig:timing-resolution-pad0} right panel. The fits that are represented as the red dashed lines of Fig. \ref{fig:timing-resolution-pad0} panels follow the Eq. \ref{eq:double_expo}. 


\begin{figure}[hbt!]
\centering 
\includegraphics[width=0.8\textwidth]{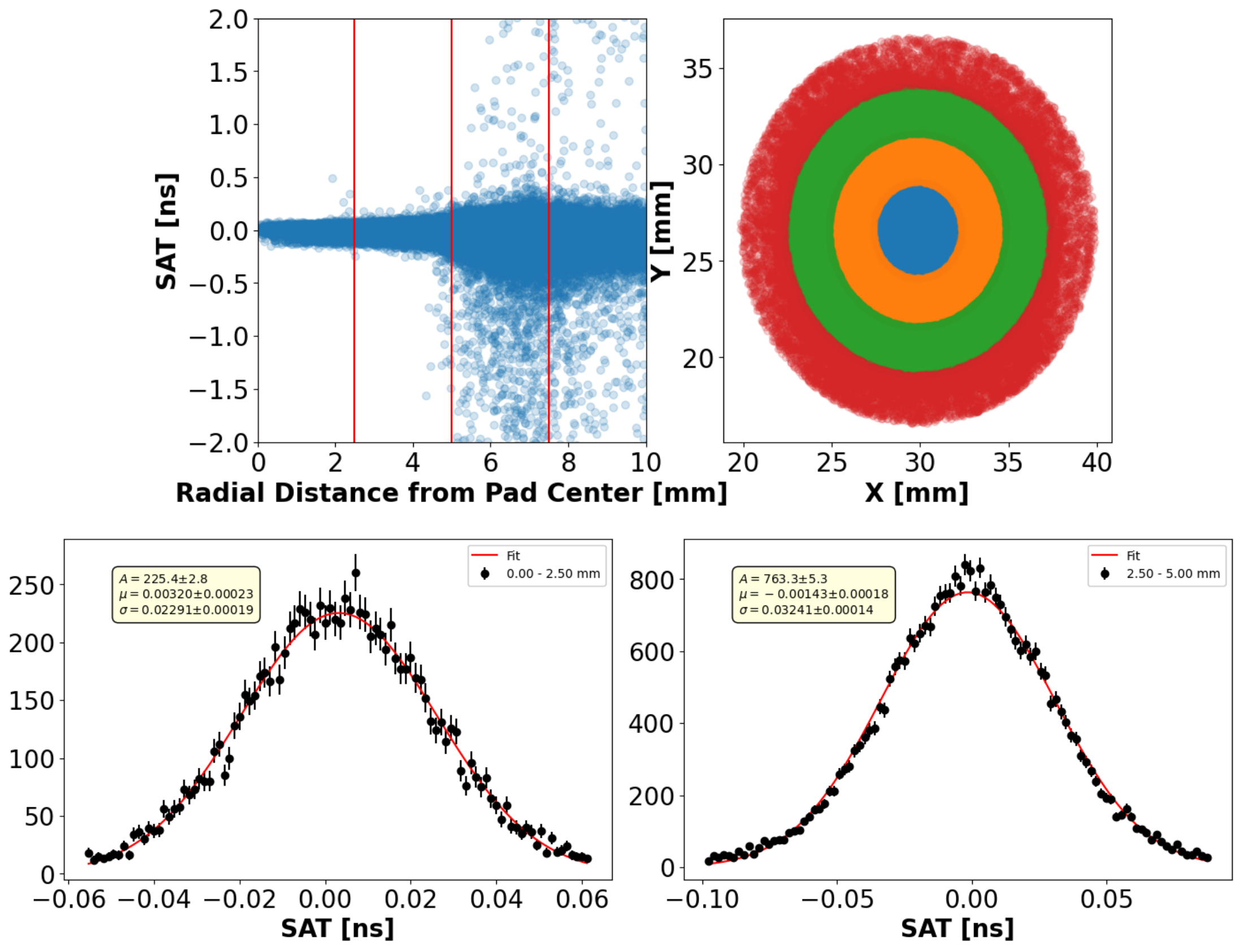}
\caption{(Top) 2D graph of the corrected SAT distribution, divided into four rings: [0, 2.5] \,mm ,  [2.5, 5] \,mm ,  [5, 7.5] \,mm and [7.5, 10] \,mm. (Bottom) Corrected timing resolution in the respective concentric rings, [0, 2.5]\,mm, [2.5, 5]\,mm, red lines correspond to Gaussian fits.}
\label{fig:sat-ring-scan}
\end{figure}

After the application of all quality and geometrical cuts, the corrected SAT and timing resolution were examined for distinct regions of each pad, defined by concentric radial zones with respect to the reconstructed pad center. Four radial regions were considered: 0–2.5\,mm, 2.5–5\,mm, 5–7.5\,mm, and 7.5–10\,mm. The results demonstrated that central regions exhibit superior timing performance. For events reconstructed within $r<2.5\,mm$ from the pad center, the Gaussian width of the time-walk corrected SAT distribution is $\sigma = 22.9 \pm 0.2$\,ps, as shown in Fig.~\ref{fig:sat-ring-scan} bottom-left. This value represents the intrinsic single-pad timing resolution for fully contained charge clusters and should be distinguished from the charge-dependent resolution shown in Fig. \ref{fig:timing-resolution-pad0}. It is worth to mention that the timing resolution gradually degrades toward the outer regions of the pad, as can be seen in the bottom right panel on the same figure.

\begin{figure}[hbt!]
\centering 
\includegraphics[width=0.5\textwidth]{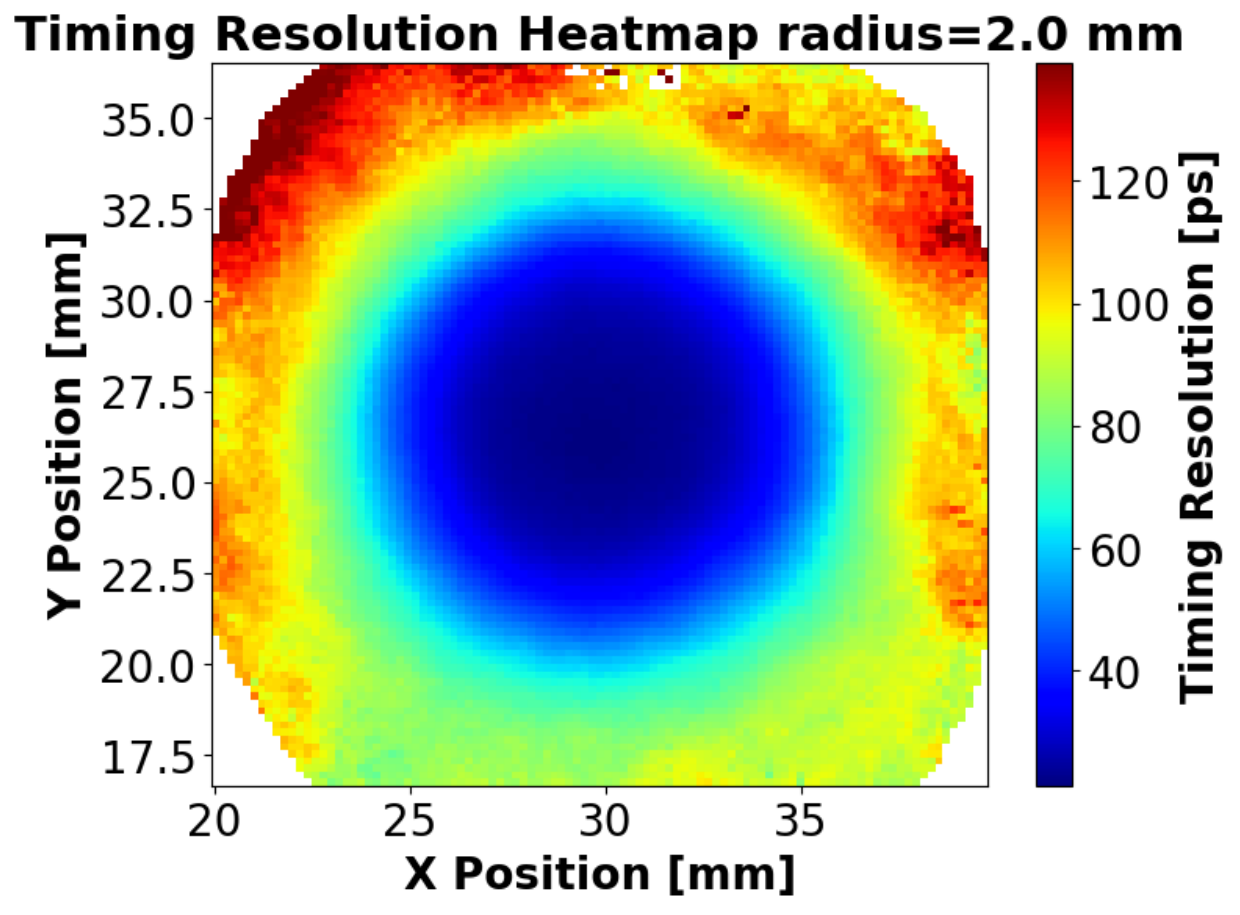}
\caption{2D circular scan producing timing heatmap for the pad-0 region.}
\label{fig:timing-sat-heatmaps}
\end{figure}

To further investigate the spatial uniformity of the timing resolution across the pad surface, an independent two-dimensional circular scan is performed. A regular Cartesian grid of $101 \times 101$ positions with a pitch of 0.2\,mm is defined over a $20 \times 20$\,mm$^2$ region centred on the pad. At each grid point, all events whose extrapolated track position falls within a circle of radius 2\,mm are collected, and a Gaussian is fitted to their corrected SAT distribution to extract the local timing resolution. Because the grid pitch (0.2\,mm) is much smaller than the circle radius (2\,mm), adjacent circles share the large majority of their events, ensuring smooth spatial coverage across the pad. The Gaussian $\sigma$ at each grid point is assigned to the corresponding pixel, and the resulting $101 \times 101$ array is displayed as a heatmap in Fig.~\ref{fig:timing-sat-heatmaps}. The timing resolution remains consistently below 30\,ps over the full 10\,mm active area of the pad. 

\begin{figure}[hbt!]
\centering 
\includegraphics[width=0.6\textwidth]{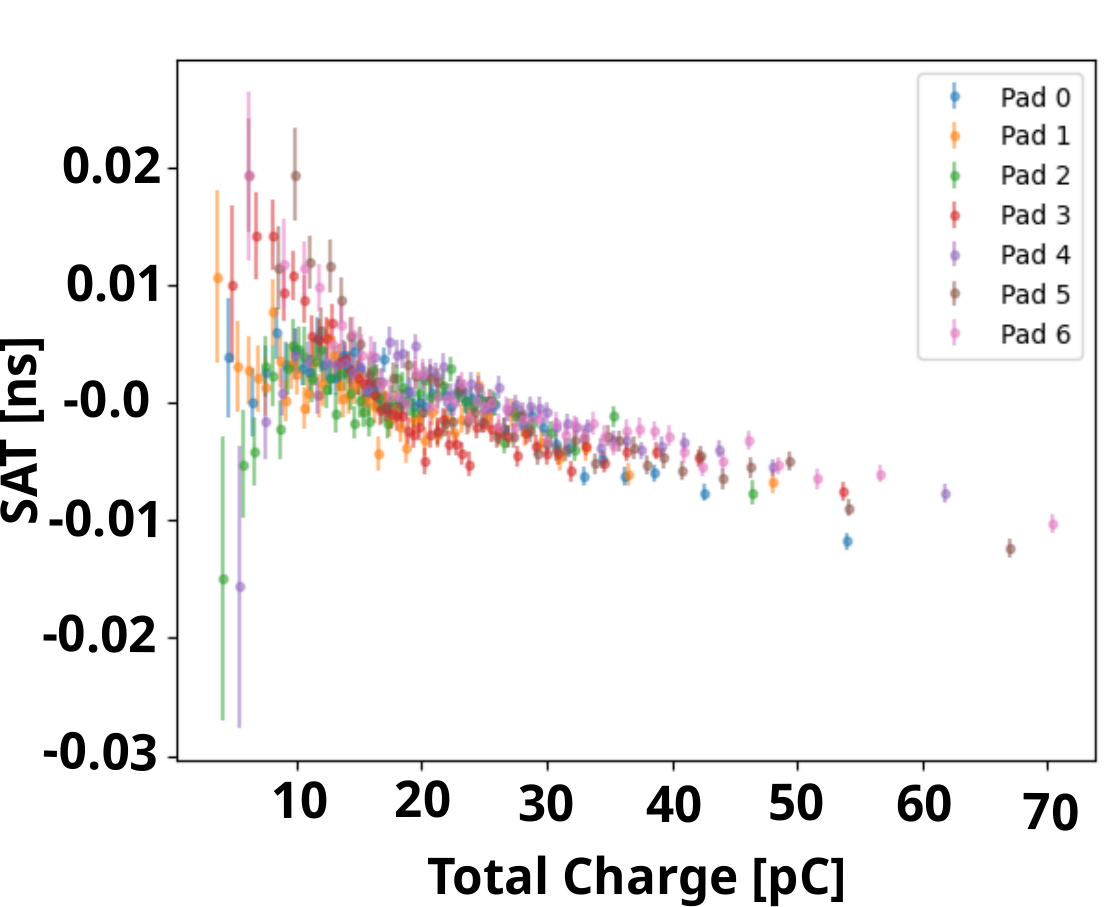}
\caption{Dependence of SAT versus total charge for all the pads of the 10\,M$\mathrm{\Omega}/\square$ prototype. The SAT values are not corrected for time-walk; just the mean values have been subtracted for better visualization.}
\label{fig:combined-sat-all-pads}
\end{figure}

Up to this point, the analysis has focused on the timing performance of individual pads, where the timing information was derived from a single pad per event. This approach is adequate when the impact point of a minimum ionizing particle (MIP) lies close to the pad center, such that the Cherenkov cone is contained entirely within that pad and most of the avalanching electrons are collected there. However, for a large-area detector, the impact point of most MIPs will be closer to the junction between neighboring pads than their center. In such cases, the induced signal is shared among multiple pads, each providing an independent measurement of the SAT.
For completeness, Fig.\ref{fig:combined-sat-all-pads} compares the SAT-charge parameterizations for all seven pads of the detector, illustrating the pad-to-pad uniformity of the time-walk behavior after cable-delay corrections.

\section{Combined Timing and Spatial Information of the Full Detector}\label{sec:combined-analysis}

To develop a truly stand-alone detector capable of precise timing over large areas, the timing reconstruction must be performed independently of any external spatial information. Relying on an external tracking system would contradict this objective, as it would introduce external spatial constraints.
The analysis conducted so far relies exclusively on the total waveform charge measured by each PICOSEC-Micromegas pad, showing that the timing resolution can be fully parameterized as a function of the collected charge.
Before proceeding further and setting aside the use of tracking data, it is essential to assess the detector’s intrinsic spatial reconstruction capability. In the following section, the external GEM telescope is used only as a reference system to evaluate the spatial resolution achievable with charge sharing among neighboring pads. 

\subsection*{First Indications of Spatial Resolution Studies}

The spatial resolution of the 7-pad prototype is therefore evaluated using a method that exploits the charge-sharing pattern among neighboring pads to reconstruct the impact position of the incident particles.

\added{Here, $(\bar{x}, \bar{y})$ are the reconstructed track impact coordinates provided by the GEM tracking telescope.}

\begin{figure}[hbt!]
\centering 
\includegraphics[width=0.8\textwidth]{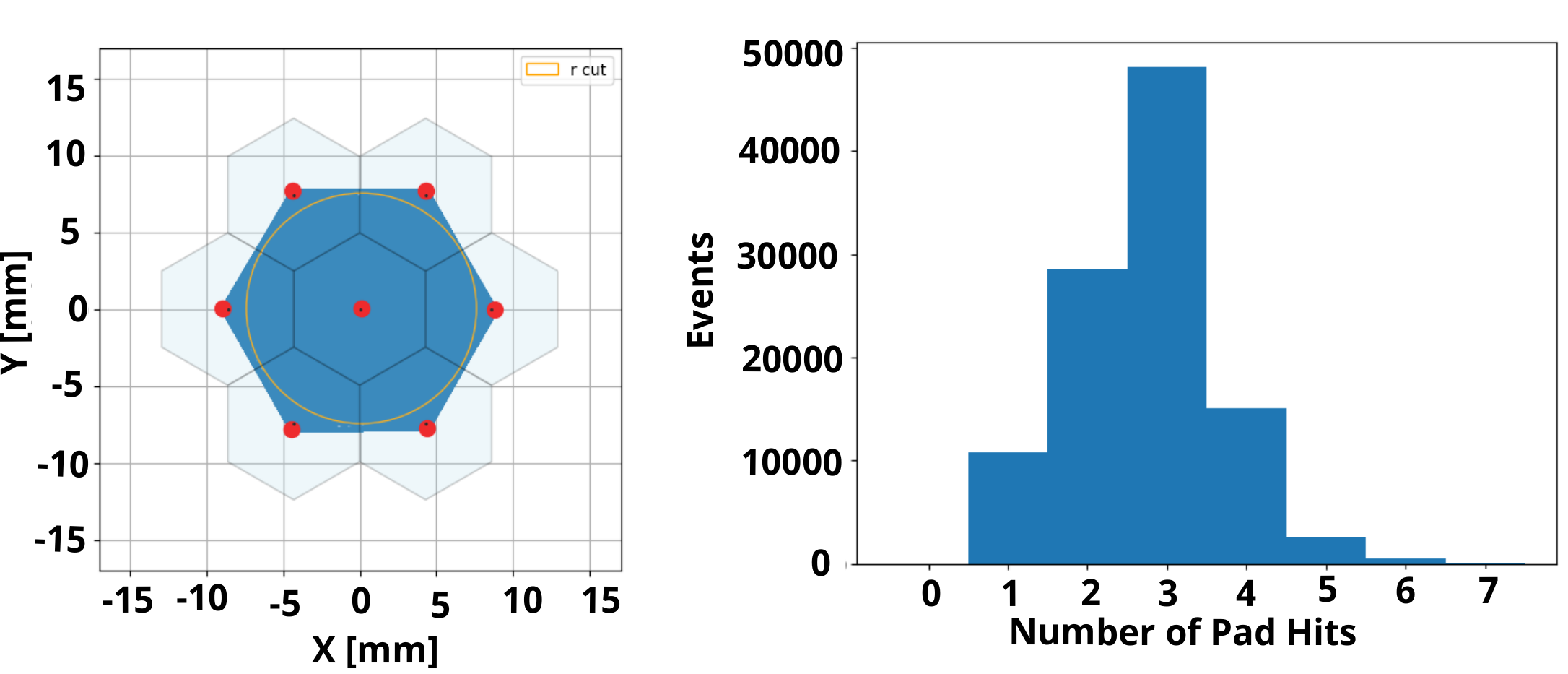}
\caption{(Left) The yellow circle represents the fiducial region used in the spatial-resolution analysis. Black markers indicate the nominal pad centers from the detector design, while the red circles show the reconstructed pad centers obtained from the alignment procedure. (Right) Histogram of $N_{hits}$ across all the events, with median at three active pads.}
\label{fig:spatial-resolution-combined}
\end{figure}

A fiducial selection is applied to ensure that reconstructed hits lie within the central region of the detector, reducing edge effects that may distort charge interpolation. The selected region corresponds to a radial distance of $ r_{center} < 7.5\,\mathrm{mm} $ from the detector’s geometric center, as shown in Fig.\ref{fig:spatial-resolution-combined}.

To reconstruct the hit position, we perform a charge-weighted interpolation over the pad centers:

\begin{align}
    x_{reco}&= \frac{\sum_i Q_i x_i}{\sum_i Q_i}, \\
    y_{reco}&= \frac{\sum_i Q_i y_i}{\sum_i Q_i}
\end{align}

where $Q_i$ is the total charge recorded on pad $i$, and $(x_i, y_i)$ are the pad center coordinates. Pads with no signal contribute zero weight.

\begin{figure}[hbt!]
\centering 
\includegraphics[width=0.8\textwidth]{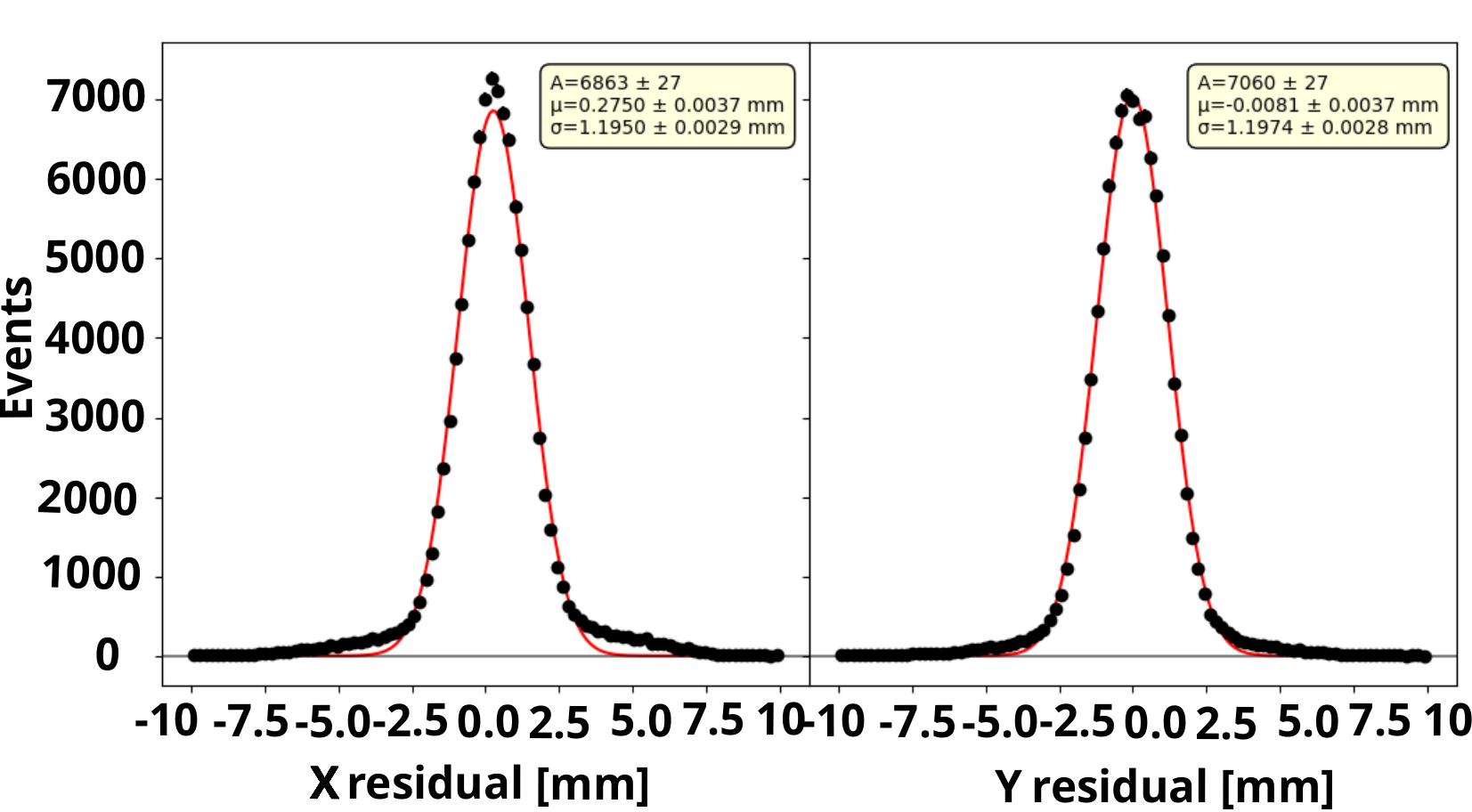}
\caption{Residual distributions in X and Y between the charge-weighted reconstructed positions and the reference track positions. The red curves show single-Gaussian fits to the core of the distributions, yielding core spatial resolutions of $\sigma_X = 1.195 \pm 0.003$\,mm and $\sigma_Y = 1.197 \pm 0.003$\,mm. The quoted uncertainty corresponds to the statistical fit error. The overall residual distributions are a superposition of events with different pad multiplicities (from 1 to 7 pads contributing), with 3-pad events dominating the core, while other multiplicity classes contribute to the broader tails.}
\label{fig:spatial-resolution-residuals}
\end{figure}

The reconstructed hit positions $(x_{reco}, y_{reco})$ are compared to the reference coordinates $(\bar{x}, \bar{y})$, and the residuals $(\delta x, \delta y)$ are computed to evaluate the spatial resolution. The core of both residual distributions is fitted with a Gaussian function, yielding core spatial resolutions of $1.195 \pm 0.003$\,mm (X) and $1.197 \pm 0.003$\,mm (Y), as shown in Fig. \ref{fig:spatial-resolution-residuals}.
The quoted uncertainty corresponds to the statistical error of the Gaussian fit, which is small due to the large event statistics. However, the residual distributions are not purely Gaussian: they represent a superposition of multiple event classes with different pad multiplicities and charge-sharing configurations, which generate non-Gaussian tails. The single-Gaussian fit is therefore used as a core spatial resolution estimate, while the tails reflect variations in charge sharing and event topology (spread from 1 to 7 pads) rather than intrinsic detector resolution. The influence of these non-Gaussian components is considered as a source of systematic uncertainty in the resolution definition.

\begin{figure}[hbt!]
\centering 
\includegraphics[width=0.8\textwidth]{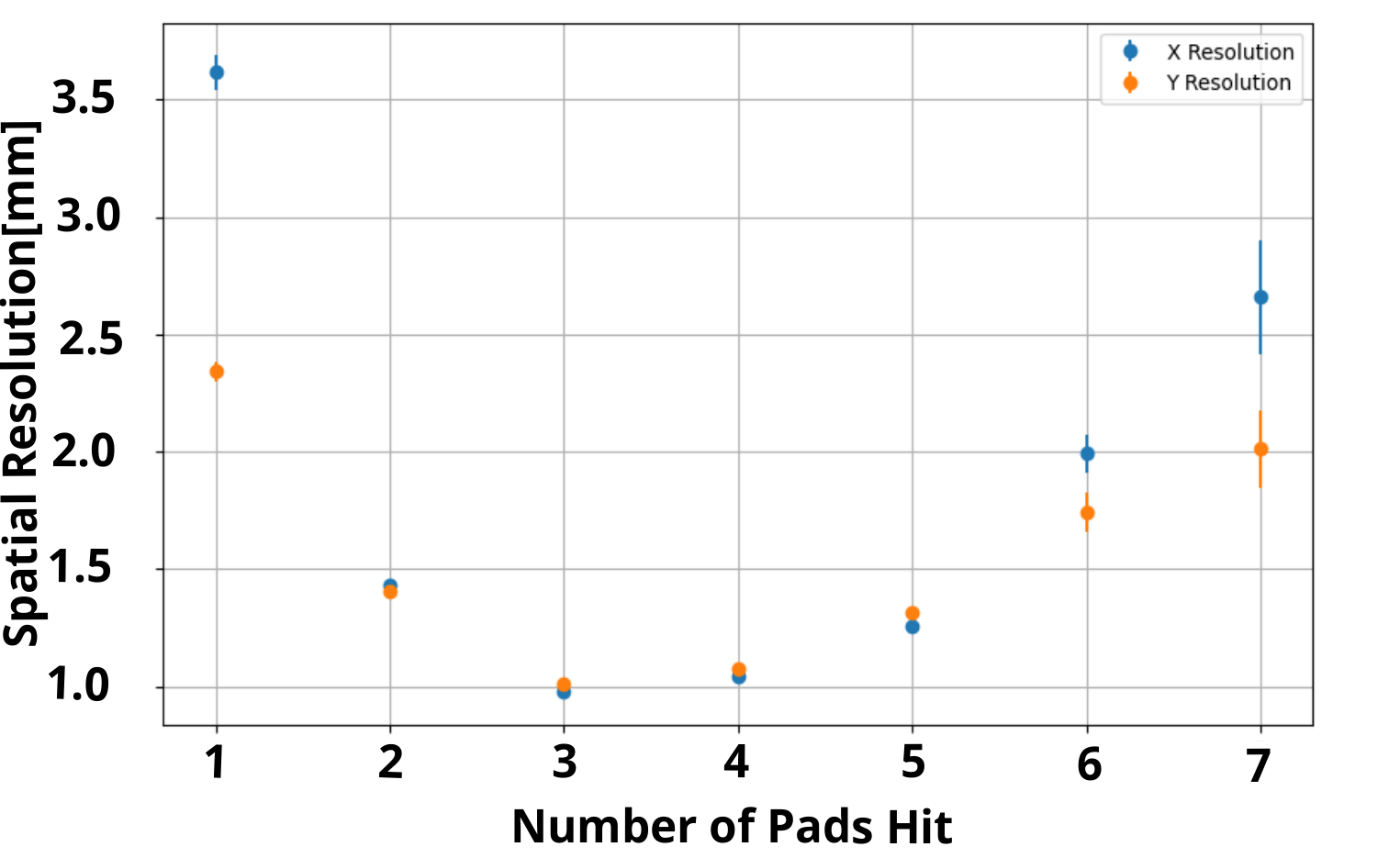}
\caption{Spatial resolution as a function of the number of pads contributing to the charge-weighted position reconstruction ($N_{\mathrm{hits}}$). The best spatial resolution is observed for events with three active pads, where charge sharing provides optimal position interpolation. The differences between the X and Y resolutions at very low and very high multiplicities arise from geometrical selection effects of the hexagonal pad layout and from the increasing influence of non-Gaussian tails in those event classes.}\label{fig:spatial-resolution-number-active-pads}
\end{figure}

Fig. \ref{fig:spatial-resolution-number-active-pads} shows that after repeating the same procedure for different pad multiplicities, the best performance is achieved for events involving three active pads, balancing information from charge sharing with minimal charge dilution effects. The x-y differences observed for the extreme multiplicities are driven by event-topology selection:conditioning on pad multiplicity selects impact-point regions with preferred orientations in the hexagonal layout, and the corresponding residual distributions are more tail-dominated than the 2-4 classes.

\subsection*{Combination of Timing Information Across the Detector Area}

A key objective of this study is to maintain the independence of the applied timing analysis from spatial information, even in events where the Cherenkov cone is shared among multiple pads. In such cases, several pads are activated simultaneously, each recording an independent measurement of the SAT. Throughout this analysis, we refer to these cases as signal-sharing events. Each pad’s timing measurement is assumed to carry independent information on the MIP arrival time, enabling a purely timing-based combination of data across the detector plane, following the assumptions made in previous studies \cite{AUNE2021165076}.

To determine the optimal method for merging timing information from neighboring pads, three combination algorithms were developed and tested. Each method is based on a general weighted-average estimator of the form:

\begin{equation}
    \hat{t_{\mathrm{comb}}} = 
    \frac{\sum_{k=1}^{K} (t^k - S(q_{\mathrm{tot}}^k)) \cdot w_k}
         {\sum_{k=1}^{K} w_k}
\end{equation}

where $K$ is the number of activated pads, $t^k$ the measured SAT of pad $k$, $S(q_{\mathrm{tot}}^k)$ the time-walk correction parameterization as a function of the total charge $q_{\mathrm{tot}}^k$, for each pad individually, as seen in Fig.\ref{fig:combined-sat-all-pads} and $w_k$ the chosen weighting factor.

The first and simplest approach selects, for each event, the pad that recorded the highest total charge and uses its corrected timing as the event timestamp. In this case, all weights are set to zero except for the pad with the maximum charge, for which $w_k=1$. This method is computationally efficient and serves as a performance baseline.

The second approach extends the idea by weighting each pad’s contribution according to its recorded total charge:

\begin{equation}
    \hat{t_{\mathrm{comb}}} = 
    \frac{\sum_{k=1}^{K} (t^k - S(q_{\mathrm{tot}}^k)) \cdot q_{\mathrm{tot}}^k}
         {\sum_{k=1}^{K} q_{\mathrm{tot}}^k}
\end{equation}

This charge-weighted average improves the representation of shared events, where multiple pads collect significant fractions of the same Cherenkov signal.

Finally, to further optimize the combination, a resolution-weighted approach is introduced. Here, each pad’s contribution is weighted by the inverse of its expected timing variance, parameterized as a function of charge:

\begin{equation}
    \hat{t_{\mathrm{comb}}} =
    \frac{\sum_{k=1}^{K} (t^k - S(q_{\mathrm{tot}}^k)) \cdot \frac{1}{R^2(q_{\mathrm{tot}}^k)}}
         {\sum_{k=1}^{K} \frac{1}{R^2(q_{\mathrm{tot}}^k)}}
\end{equation}

Here $R(q_{\mathrm{tot}}^k)$ is the timing resolution of pad $k$ as a function of its collected charge, obtained from the charge-binned Gaussian fit to the corrected SAT distribution described in Section~\ref{sec:timing-calibration} and shown in Fig.~\ref{fig:timing-resolution-pad0}. Specifically, $R(q)$ is evaluated using the double-exponential parameterisation of Eq.~\ref{eq:double_expo} fitted to the per-charge-bin Gaussian widths. This formulation gives higher influence to pads with intrinsically better timing resolution. To ensure numerical stability and suppress noise-only contributions, pads with $q_{tot}^k <2$\,pC are excluded from the sums; for selected events at least one pad remains above this threshold, so no event-level inefficiency is introduced.

\begin{figure}[hbt!]
\centering 
\includegraphics[width=1.0\textwidth]{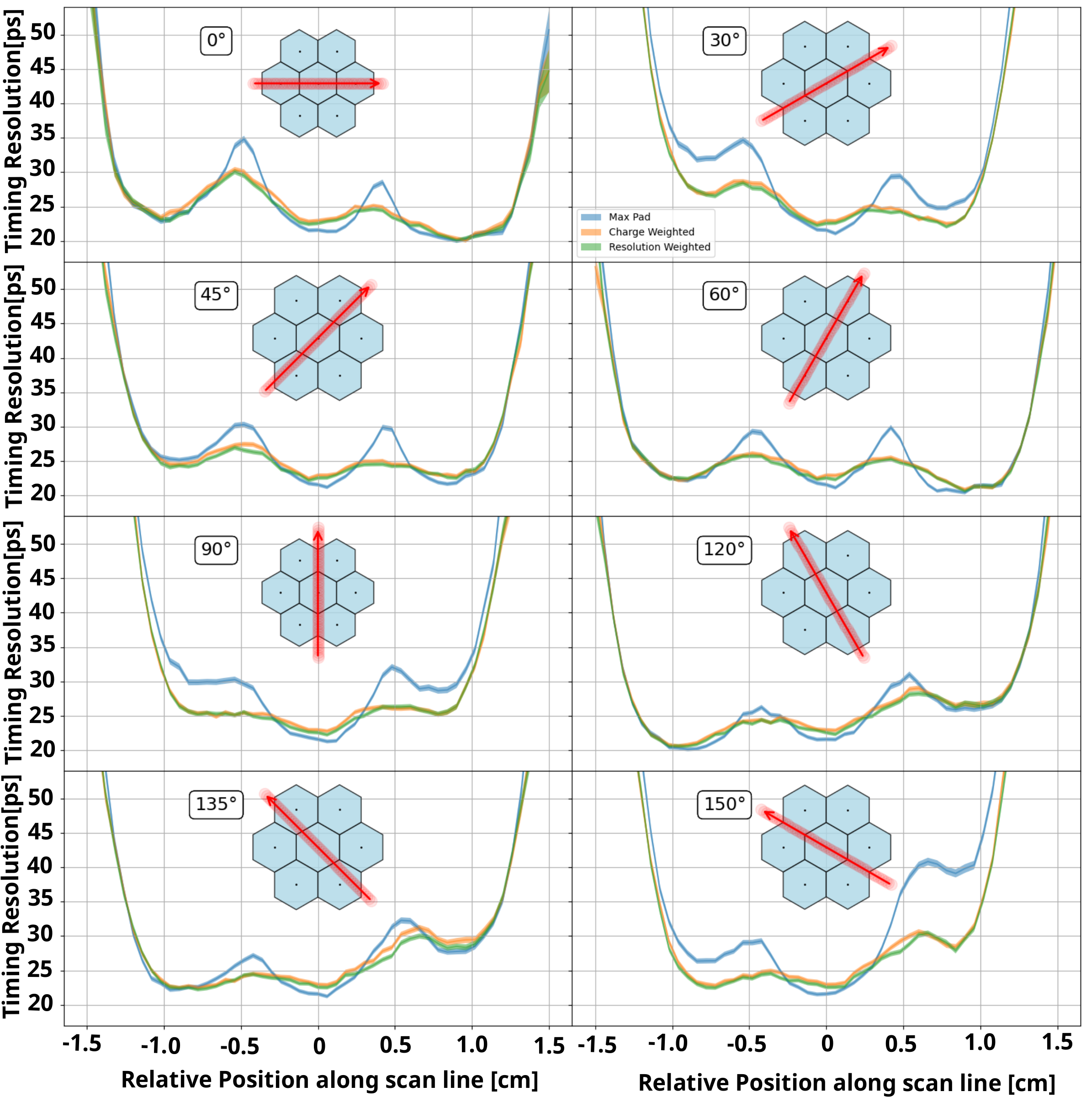}
\caption{Comparison of different weighted average algorithms for combining timing information over the total detector area. With blue, the line corresponds to the maximum pad weighted average, with orange as the charge weighted average, and green as the resolution weighted average. All the plots correspond to eight different observation paths of: 0\textdegree, 30\textdegree, 45\textdegree, 60\textdegree, 90\textdegree, 120\textdegree, 135\textdegree, and 150\textdegree.}
\label{fig:10MO-SAT-vs-angle-combined}
\end{figure}

The three algorithms are compared by scanning the detector along eight linear strips at different angles: 0°, 30°, 45°, 60°, 90°, 120°, 135°, and 150°. Each strip is a sequence of sampling points equally spaced along the strip direction, spanning $\pm 2\,r_{\mathrm{outer}}$ ($\pm$10\,mm) from the detector centre. The direction of each strip is indicated by the red arrow in the corresponding subplot of Fig.~\ref{fig:10MO-SAT-vs-angle-combined}. At each sampling point, all events whose extrapolated track position falls within a circle of 2\,mm radius are collected, and the three timing combination algorithms are evaluated on those events. Scanning at multiple angles probes the spatial uniformity of the combined timing response across the full pad area and across pad boundaries, where charge is shared between neighbouring pads.

Using this systematic scan, the combined SAT values obtained from each algorithm were compared over the whole detector area, producing the results shown in Fig.~\ref{fig:10MO-SAT-vs-angle-combined}. The ``maximum-charge'' approach (blue curve) performs best near pad centers, while the ``resolution-weighted'' method (green curve) provides superior response in inter-pad regions, where signals are shared across neighboring channels. The ``charge-weighted'' approach (orange curve) performs consistently worse than the resolution-weighted method across all regions, including pad centers, and does not offer a clear advantage over either of the other two approaches.

Overall, the results demonstrate that combining timing data from multiple pads without relying on external spatial information is both feasible and effective. The choice of algorithm may depend on the experiment’s specific performance requirements and computational constraints. In practice, the resolution-weighted combination offers the most balanced and robust approach for accurate time reconstruction across shared regions.

\begin{figure}[hbt!]
\centering 
\includegraphics[width=1.0\textwidth]{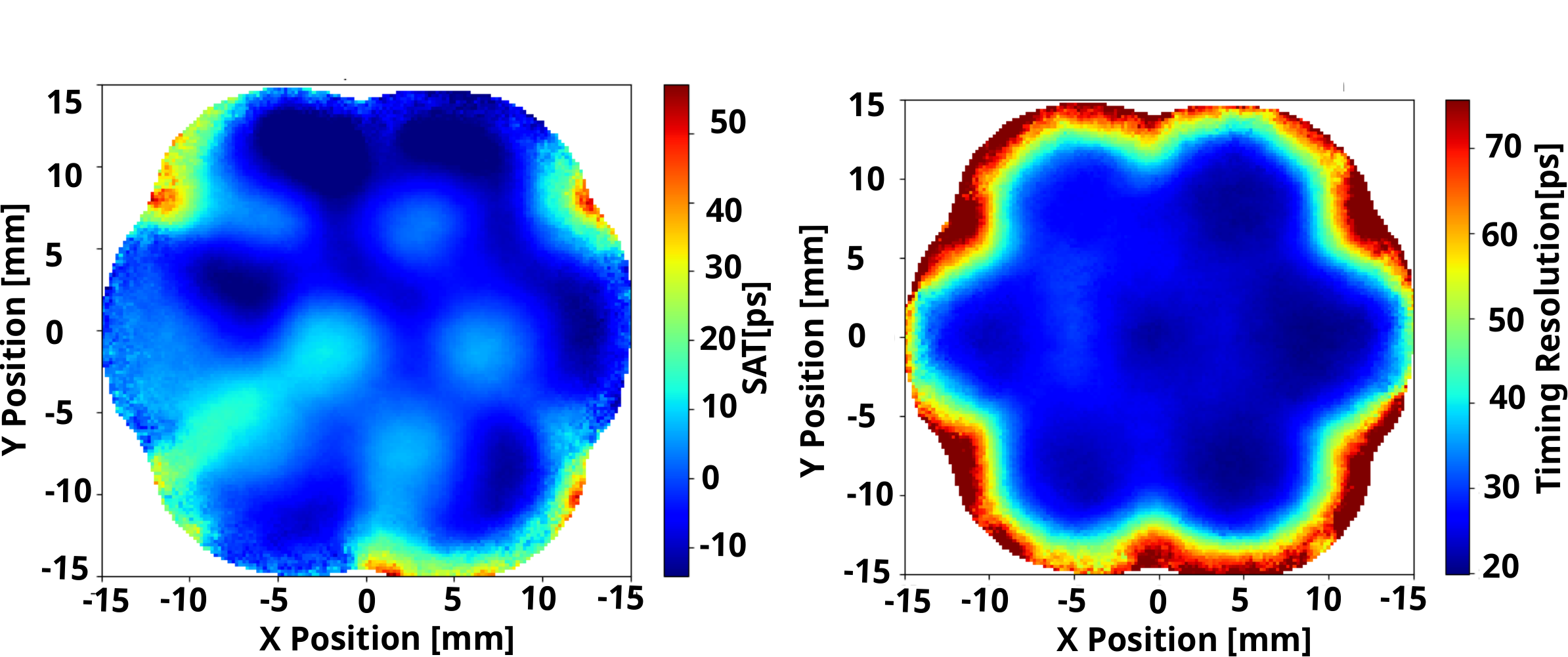}
\caption{Combined corrected SAT and timing resolution 2D circular scan of the whole detector area.}
\label{fig:timing-sat-combined}
\end{figure}

The resulting combined SAT and timing resolution maps are shown in Fig.\ref{fig:timing-sat-combined}. The SAT map reveals minor systematic spatial variations across the detector surface. These are attributed to a residual tilt between the photocathode and the readout plane, a consequence either of the photocathode support being fixed with only three peripheral screws, or of the non-planarity of the readout board, which does not guarantee sub-10\,\textmu m planarity uniformity. As demonstrated in \cite{AUNE2021165076}, such geometric imperfections introduce non-uniformities in the drift electric field that produce position-dependent shifts in the SAT. The timing resolution map, by contrast, confirms uniform performance across the detector, achieving an average resolution of approximately 30--32\,ps even at pad junctions.

\subsection*{Focusing in Charge Sharing}

The final step of the analysis focuses on the detector’s performance in regions where charge sharing occurs between neighboring pads. This study provides additional validation of the timing combination algorithms discussed above, ensuring their robustness in areas where the Cherenkov cone overlaps multiple readout pads.

\begin{figure}[hbt!]
\centering 
\includegraphics[width=1.0\textwidth]{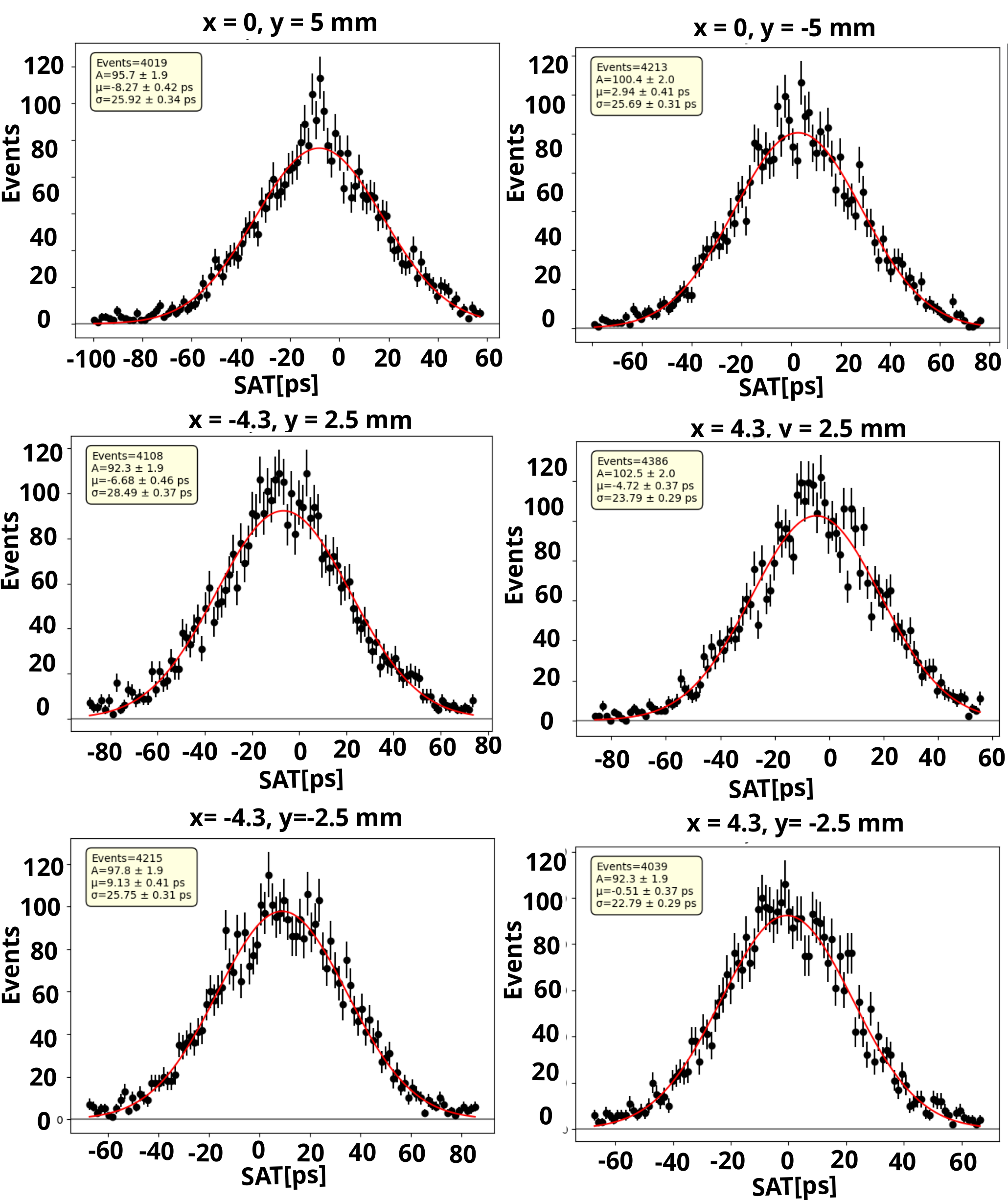}
\caption{Combined corrected timing resolution for the six common corners between the central pad (pad-0) and its neighbours. Each subplot is labelled with the $(x, y)$ coordinates (in mm) of the corresponding corner of pad-0 in the local detector frame, consistent with the reference frame shown in Fig.~\ref{fig:spatial-resolution-combined}. A consistent timing resolution below 28\,ps is observed at all six corners, validating the robustness of the resolution-weighted timing combination in shared-charge regions.}
\label{fig:sharing-points}
\end{figure}

To specifically investigate these shared-signal regions, we selected events where the reconstructed hit position lies within the six common corners between the central pad and its adjacent neighbors. In these areas, the Cherenkov cone is typically distributed across an average of three pads, resulting in comparable charge fractions being recorded by multiple channels. For each shared region, the combined timing was computed using the resolution-weighted algorithm described previously, and the resulting SAT distributions were fitted with Gaussian functions to extract the timing resolution. The results, shown in Fig.~\ref{fig:sharing-points}, demonstrate that even in shared regions, a timing resolution better than 28\,ps can be achieved, confirming the stability of the combination method across pad boundaries. This value is better than the $\sim$30--32\,ps observed in the 2D heatmap scan because it corresponds to events specifically selected at the six corner positions where charge is shared among approximately three pads, providing more favourable conditions for the resolution-weighted combination.

To complement the timing performance studies, we also derived the local detection efficiency and event statistics over the entire detector area. The 2D maps were obtained using the same Cartesian grid scan procedure described previously, but with a reduced circle radius of 1.5\,mm (compared to 2\,mm used for the timing scan) to achieve finer spatial detail in the efficiency map. At each point, the local detection efficiency $\epsilon(x_i, y_j)$ is defined as:

\begin{equation}
    \epsilon (x_i, y_j) = \frac{N_{\mathrm{valid}}(x_i,y_j)}{N_{\mathrm{total}}(x_i,y_j)}
\end{equation}

where $N_{\mathrm{valid}}(x_i,y_j)$ is the number of events satisfying the detection condition—defined as at least one pad recording a signal with a time difference smaller than 1\,ns relative to the reference—and $N_{\mathrm{total}}(x_i,y_j)$ is the total number of events within the corresponding circular region used in the scan. 

\begin{figure}[hbt!]
\centering
\includegraphics[width=0.6\textwidth]{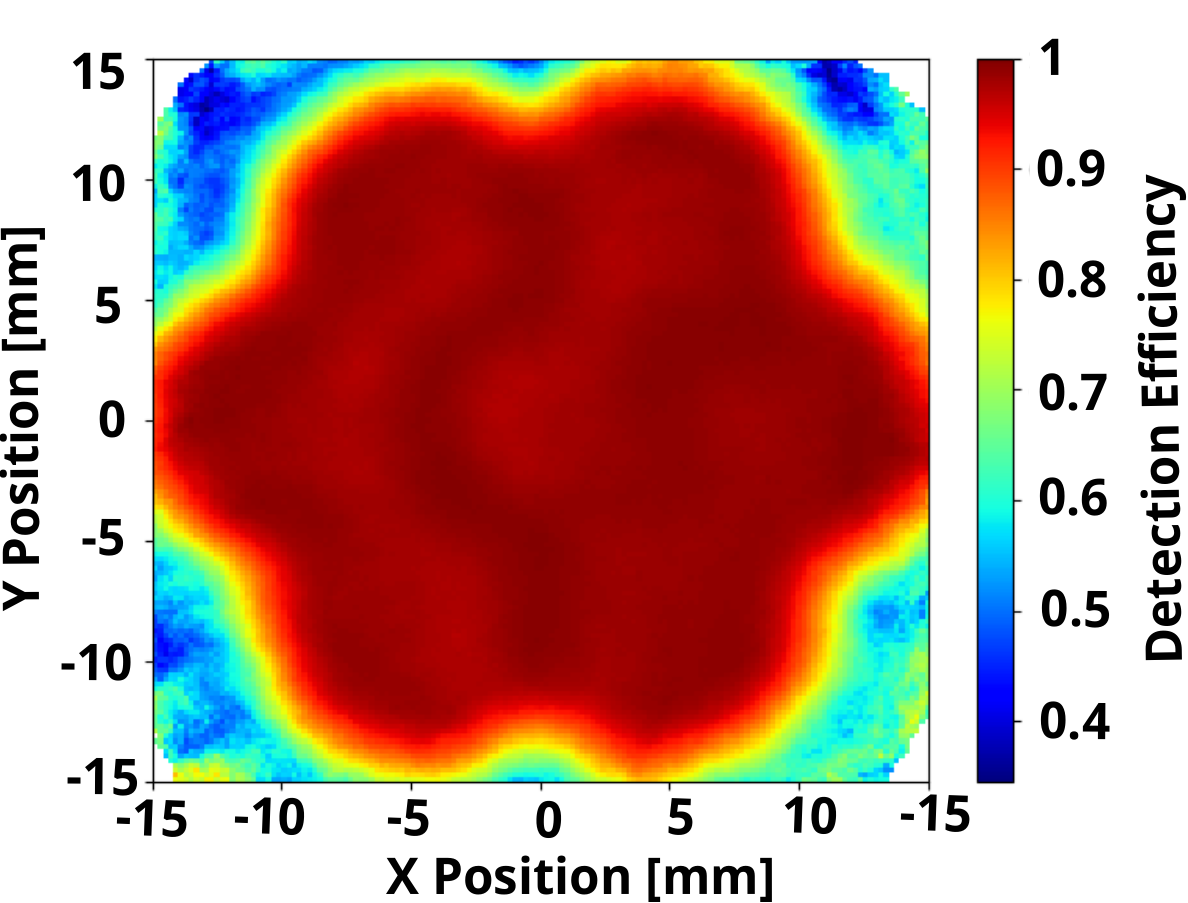}
\caption{Efficiency map over the full detector area. The non-zero efficiency extending $\sim$3\,mm beyond the active pad boundaries reflects the finite Cherenkov cone size and lateral charge spreading through the resistive layer.}
\label{fig:efficiency-10MO}
\end{figure}

The resulting map, shown in Fig.~\ref{fig:efficiency-10MO}, illustrates both the spatial uniformity of the detector response and the event distribution during the scanning run. The high efficiency observed across the active area further supports the uniformity of charge collection and the robustness of the signal reconstruction methods employed. A non-zero efficiency is observed in a region extending approximately 3\,mm beyond the geometrical boundary of the active pads. This arises from two combined effects: first, the finite transverse size of the Cherenkov photon cone means that tracks reconstructed just outside the active area can still illuminate the edge pads and produce a detectable signal; second, the resistive DLC layer spreads the induced charge laterally, allowing signals from hits in the immediate vicinity of the pad boundary to be collected by the nearest active pad. This extended response is therefore a physical feature of the detector design rather than a reconstruction artefact.

In contrast to the timing-combination studies, no minimum charge threshold is applied here, so that the efficiency reflects the intrinsic detection capability rather than an algorithmic selection.

\section{Conclusion and Discussion}\label{sec:conclusion}

This work presented a detailed performance study of a resistive PICOSEC Micromegas detector prototype tested at the CERN SPS H4 beam line facility using 150\,GeV/c muons. The primary focus was to establish and validate a comprehensive analysis framework for timing and spatial performance evaluation, using a 10\,M$\mathrm{\Omega}/\square$ resistive-plane prototype as the reference configuration. This detector exhibited the most uniform and stable response, among all tested prototypes, allowing a robust and systematic study of charge sharing, time-walk correction, timing combination strategies and spatial reconstruction methods. 

In parallel, the analysis framework was applied to a 200\,k$\mathrm{\Omega}/\square$ prototype, tested under identical conditions, to evaluate the impact of reduced resistivity on charge transport, timing resolution, and spatial reconstruction.

These studies were motivated by the ENUBET project requirements, which emphasize detector stability under moderate-rate conditions (approximately 15\,kHz over a 10\,cm diameter muon beam, representative of the SPS H4 beam conditions used here), efficient spark quenching, and minimal recovery time to guarantee long-term operation \cite{kallitsopoulou:tel-05267379}.

The 10\,M$\mathrm{\Omega}/\square$ resistive-plane prototype demonstrated the best overall balance between timing and spatial performance, achieving a timing resolution of $22.9 \pm 0.2$\,ps in single-pad events and core spatial resolutions of $1.195 \pm 0.003$\,mm (X) and $1.197 \pm 0.003$\,mm (Y). Charge sharing occurred across an average of three pads, and the detector exhibited uniform timing performance throughout the active area. In regions where signals were shared between neighboring pads, the weighted averaging algorithm — which combines the individual pad SAT using timing-resolution-based weights — recovered a combined timing resolution below 28\,ps.

The comparison between the two resistive-plane prototypes highlights the strong influence of resistivity on detector performance. Resistivity plays a key role in shaping charge transport: higher resistivity localises the induced charge, reducing lateral spread and improving pad centre estimation, while lower resistivity enhances charge diffusion, leading to broader signal distributions and small systematic shifts in reconstructed positions.

The prototype featuring a 200\,k$\mathrm{\Omega/\square}$ resistive layer displayed a small but statistically consistent shift in reconstructed pad centers relative to their geometric locations. This offset originates from the enhanced charge spread facilitated by the lower resistivity. Nevertheless, the reconstruction algorithm, which identifies the pad with the highest collected charge as the likely impact center, remained robust. This configuration reached a timing resolution of $31.6 \pm 0.3$\,ps and core spatial resolutions of $1.374 \pm 0.004$\,mm (X) and $1.132 \pm 0.003$\,mm (Y), with signals typically shared across four pads. When applying the same weighted averaging combination, the resulting timing resolution was approximately 30\,ps. 

The minor systematic spatial variations observed in the combined SAT heatmap (Fig.\,\ref{fig:timing-sat-combined}) are attributed to photocathode non-parallelism with the readout PCB and to residual PCB planarity imperfections. As established in \cite{AUNE2021165076}, such geometric deviations produce non-uniform drift electric fields that translate into position-dependent SAT shifts across the detector surface. The current photocathode support system — relying on only three peripheral screws to hold the crystal — does not guarantee the sub-10\,\textmu m planarity uniformity required for optimal timing homogeneity; deviations exceeding 20\,\textmu m lead to measurable local variations in the drift gap thickness and electric field. These observations motivate the development of improved mechanical designs in next-generation detectors, where enhanced planarity control is a primary engineering objective \cite{kallitsopoulou2026designperformance96channelresistive}.

The insights gained from these studies directly guided the design choices for the next-generation 96-pad detector prototype \cite{kallitsopoulou2026designperformance96channelresistive}. The 10\,M$\mathrm{\Omega}/\square$ resistive-plane architecture was selected as the optimal configuration, offering the best balance of timing performance, spatial resolution, and mechanical robustness. A summary of the key performance metrics for the two resistive-plane prototypes characterised in this work is given in Tab.\ref{tab:prototype_summary}. Ongoing work focuses on enhancing mechanical stability and ensuring sub-10\,\textmu m planarity, to fully realise the potential of this resistive Micromegas technology for precision timing applications within the ENUBET experimental framework.

\begin{table}[htb!]
\centering
\caption{Summary of performance metrics for resistive PICOSEC Micromegas prototypes. Both prototypes use the same hexagonal pad geometry; the higher average pad sharing observed for the 200\,k$\Omega/\square$ prototype (4 pads vs.\ 3 pads) is a consequence of the enhanced lateral charge spreading through the lower-resistivity DLC layer, which distributes the induced charge over a broader area.}
\resizebox{\linewidth}{!}{%
\begin{tabular}{|c|c|c|c|c|}
\hline
Resistive Layer Architecture & Timing[\diameter8\,mm](ps) & Spatial X/Y (mm) & Pad Sharing & Charge Sharing Timing (ps) \\
\hline
10\,M$\Omega/\square$ Resistive plane & $22.9 \pm 0.2$ & $1.195 \pm 0.003$ / $1.197 \pm 0.003$ & 3 pads & $<$28 \\
\hline
200\,k$\Omega/\square$ Resistive plane & 31.6 $\pm$ 0.3 & $1.374 \pm 0.004$ / $1.132 \pm 0.003$ & 4 pads & $<$30 \\
\hline
\end{tabular}%
}
\label{tab:prototype_summary}
\end{table}

\acknowledgments
The authors gratefully acknowledge the support of the French National Research Agency (ANR) through the project \textit{“Development of a PICOSEC-Micromegas detector for ENUBET – PIMENT”}(ANR-21-CE31-0027). The authors also acknowledge the support of the RD51 collaboration in the framework of RD51 common projects. We acknowledge the financial support of the CrossDisciplinary Program on Instrumentation and Detection of CEA, the French Alternative Energies and Atomic Energy Commission. This work was partially supported by the European Union’s Horizon 2020 research and innovation program through the STRONG-2020 project under grant agreement No.~824093.

\bibliographystyle{JHEP}
\bibliography{main.bib}
\end{document}